# A TRANSIENT TRANSIT SIGNATURE ASSOCIATED WITH THE YOUNG STAR RIK-210

Trevor J. David[1], Erik A. Petigura[2], Lynne A. Hillenbrand[1], Ann Marie Cody[3], Andrew Collier Cameron[4], John R. Stauffer[5], B.J. Fulton[6,7], Howard T. Isaacson[8], Andrew W. Howard[1,6], Steve B. Howell[3], Mark E. Everett[9], Ji Wang[1], Björn Benneke[2], Coel Hellier[10], Richard G. West[11], Don Pollacco[11], David R. Anderson[10]

## ABSTRACT

We find transient, transit-like dimming events within the *K2* time series photometry of the young star RIK-210 in the Upper Scorpius OB association. These dimming events are variable in depth, duration, and morphology. High spatial resolution imaging revealed the star is single, and radial velocity monitoring indicated that the dimming events can not be due to an eclipsing stellar or brown dwarf companion. Archival and follow-up photometry suggest the dimming events are transient in nature. The variable morphology of the dimming events suggests they are not due to a single, spherical body. The ingress of each dimming event is always shallower than egress, as one would expect for an orbiting body with a leading tail. The dimming events are periodic and synchronous with the stellar rotation. However, we argue it is unlikely the dimming events could be attributed to anything on the stellar surface based on the observed depths and durations. Variable obscuration by a protoplanetary disk is unlikely on the basis that the star is not actively accreting and lacks the infrared excess associated with an inner disk. Rather, we explore the possibilities that the dimming events are due to magnetospheric clouds, a transiting protoplanet surrounded by circumplanetary dust and debris, eccentric orbiting bodies undergoing periodic tidal disruption, or an extended field of dust or debris near the corotation radius.

## 1. INTRODUCTION

Upper Scorpius, hereafter Upper Sco, is the nearest OB association (see Preibisch & Mamajek 2008, for a review). At an age of 5–10 Myr, the association is at a critical stage of planet formation, when most protoplanetary disks have dissipated. Roughly 20% of the low-mass members of Upper Sco host a protoplanetary disk, indicating that planet formation is ongoing in the region (Luhman & Mamajek 2012).

Around young ($\lesssim$10 Myr) planet-forming stars, dimming events of several tens of percent have been explained as obscuration by a circumstellar disk (Cody et al. 2014). Indeed, in Upper Sco itself, dozens of members exhibit such dimming behavior (Ansdell et al. 2016). At the age of Upper Sco, it is also possible to find fully-formed planets at small orbital separations, as evidenced by *K2*–33 b (David et al. 2016b; Mann et al. 2016), a transiting, Neptune-sized, short-period exoplanet around a low-mass member of Upper Sco.

Transit profiles that are asymmetric and variable in depth and duration may be attributable to disintegrating planetary bodies with trailing or leading tails (Rappaport et al. 2012, 2014; Sanchis-Ojeda et al. 2015; Vanderburg et al. 2015), swarms of rocky or cometary debris (Boyajian et al. 2016), circumplanetary rings (Mamajek et al. 2012), or a precessing planet transiting a star with non-uniform surface brightness (Barnes et al. 2013). While this phenomenon has been documented around mature stars and stellar remnants, due to e.g. tidal disruption or photoevaporation, *periodic* examples have not been previously observed around a young star lacking a protoplanetary disk. Here we present evidence of a transient transit-like signature, possibly due to a transiting cloud or an enshrouded protoplanet, around the young star RIK-210.

In §2 we describe existing and new information about the star. In §3 we present the *K2* data and light curve analysis. Archival and follow-up ground-based photometry is presented in §4, time series spectroscopy in §5, and high spatial resolution imaging in §6. §7 discusses the physical interpretation of the variable depth narrow flux dips in RIK-210, including stellar surface activity and several orbiting planet/debris scenarios.

## 2. RIK-210

RIK-210 (also designated 2MASS J16232454-1717270 and EPIC 205483258) was established as a low-mass member of the 5–10 Myr old Upper Scorpius OB association by Rizzuto et al. (2015). Those authors assigned a 95% membership probability on the basis of hydrogen emission and lithium absorption. Proper motions from

tjd@astro.caltech.edu
[1] Department of Astronomy, California Institute of Technology, Pasadena, CA 91125, USA
[2] Division of Geological and Planetary Sciences, California Institute of Technology, Pasadena, CA 91125, USA
[3] NASA Ames Research Center, Mountain View, California 94035, USA
[4] SUPA, School of Physics and Astronomy, University of St Andrews, St Andrews, KY16 9SS, UK
[5] Spitzer Science Center, California Institute of Technology, Pasadena, CA 91125, USA
[6] Institute for Astronomy, University of Hawaii at Manoa, Honolulu, Hawaii 96822, USA
[7] NSF Graduate Research Fellow
[8] Department of Astronomy, University of California, Berkeley, California 94720, USA
[9] National Optical Astronomy Observatory, 950 N. Cherry Ave., Tucson, AZ 85719, USA
[10] Astrophysics Group, Keele University, Staffordshire, ST5 5BG, UK
[11] Department of Physics, University of Warwick, Coventry CV4 7AL, UK

**Table 1**
System properties of RIK-210

| Parameter | Value | Reference |
|---|---|---|
| 2MASS designation | J16232454-1717270 | |
| EPIC designation | 205483258 | |
| Spectral type | M2.5 | this work |
| $d$ (pc) | $145 \pm 20$ | de Zeeuw et al. (1999) |
| $A_V$ (mag) | $0.6 \pm 0.2$ | this work |
| EW H$\alpha$ (Å) | -3.5 to -6.5 | this work |
| EW He I 5876 (Å) | -0.6 to -1.2 | this work |
| EW Li (Å) | $0.54 \pm 0.02$ | this work |
| $\log (T_{\rm eff}/{\rm K})$ | $3.54 \pm 0.01$ | this work |
| $\log L_*/L_\odot$ (dex) | $-0.70 \pm 0.08$ | this work |
| $M_*$ ($M_\odot$) | $0.53^{+0.13}_{-0.13}$ | this work |
| $R_*$ ($R_\odot$) | $1.24^{+0.12}_{-0.13}$ | this work |
| $\gamma$ (km s$^{-1}$) | $-4.63 \pm 0.07$ | this work |
| $v \sin i$ (km s$^{-1}$) | $11 \pm 1$ | this work |
| $P_{\rm rot}$ (d) | $5.670 \pm 0.004$ | this work |
| $Kp$ (mag) | 13.7 | EPIC |
| $B$ (mag) | $16.213 \pm 0.066$ | APASS DR9 |
| $V$ (mag) | $14.630 \pm 0.137$ | APASS DR9 |
| $g'$ (mag) | $15.462 \pm 0.121$ | APASS DR9 |
| $r'$ (mag) | $14.093 \pm 0.118$ | APASS DR9 |
| $i'$ (mag) | $12.862 \pm 0.099$ | APASS DR9 |
| $J$ (mag) | $10.61 \pm 0.02$ | 2MASS |
| $H$ (mag) | $9.85 \pm 0.02$ | 2MASS |
| $K$ (mag) | $9.65 \pm 0.02$ | 2MASS |
| $W1$ (mag) | $9.55 \pm 0.02$ | WISE |
| $W2$ (mag) | $9.40 \pm 0.02$ | WISE |
| $W3$ (mag) | $9.19 \pm 0.05$ | WISE |
| $W4$ (mag) | $8.75 \pm 0.45$ | WISE |

the UCAC4 catalog (Zacharias et al. 2013) are consistent with membership (de Zeeuw et al. 1999; Lodieu 2013). We confirm the youth of the star spectroscopically and through a precise determination of its systemic radial velocity.

### 2.1. *Stellar properties*

Observed and derived stellar properties for RIK-210 are reported in Table 1. For all distance-dependent parameters we assumed $d = 145 \pm 20$ pc, corresponding to the mean distance to Upper Scorpius from *Hipparcos* trigonometric parallaxes of the high-mass members and assuming an uncertainty comparable to the association's width on the sky (de Zeeuw et al. 1999; de Bruijne 1999). Our spectrum (§6) is consistent with the previously reported spectral type of M2.5 (Rizzuto et al. 2015). Assuming the spectral type and using empirical calibrations valid for young stars (Pecaut & Mamajek 2013), we derived the effective temperature, bolometric luminosity and visual extinction for RIK-210. We determined the stellar radius from the temperature, luminosity, and the Stefan-Boltzmann law.

The stellar mass was determined from the temperature and luminosity, interpolating between solar metallicity (Z=0.0152) PARSEC v1.2s pre-main sequence models (Bressan et al. 2012; Chen et al. 2014). Mass uncertainties were estimated from Monte Carlo sampling assuming normally distributed errors in $\log (T_{\rm eff})$ and $\log (L/L_\odot)$. We adopt generous mass uncertainties (the boundaries of the 5% and 95% quantiles, see Table 1) due to the incompleteness of models at low stellar masses to include important physics such as the magnetic inhibition of convection (e.g. Feiden 2016).

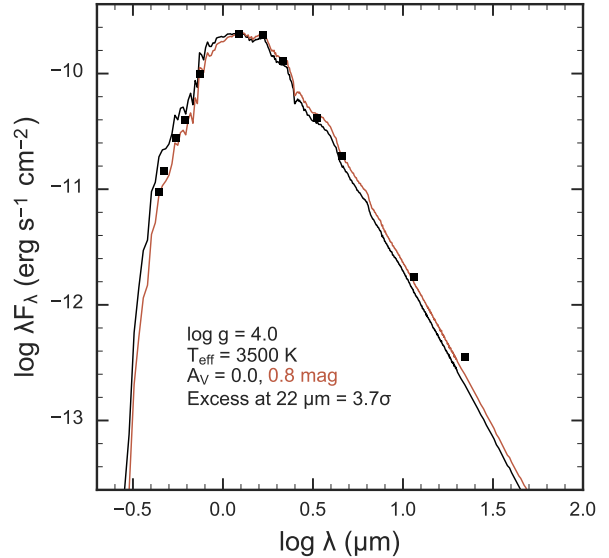

**Figure 1.** Spectral energy distribution of RIK-210 along with a NextGen stellar atmosphere model (Hauschildt et al. 1999) normalized to the J-band point. Including a small amount of reddening (black) improves the fit over the unreddened model (green).

### 2.2. *Activity and Possible Disk*

RIK-210 shows modest chromospheric emission. H$\alpha$ with a classic double-horn profile, H$\beta$, He I 5876Å, Na I D, the Ca II H&K doublet and the "infrared" triplet, and Fe I 5169 Å are all seen in emission in our spectra (§6). The line strengths are variable over our spectral time series, as discussed in detail below.

Although the star is active, the spectrum indicates that it is not currently accreting. Furthermore, the spectral energy distribution (Figure 1) shows that it does not host a primordial protoplanetary disk. However, there is marginal evidence for a 22 $\mu$m mid-infrared excess at the 40% level (detected at SNR > 3.5), based on a model atmosphere fit to available broadband catalog photometry. Typical dust masses of low-mass stars with circumstellar disks in Upper Sco are in the range of 0.2–20 M$_\oplus$ (Barenfeld et al. 2016). As RIK-210 lacks any significant circumstellar disk, we assume the amount of remaining dust in the system is likely below 0.2 M$_\oplus$.

## 3. *K2* LIGHT CURVE AND ANALYSIS

RIK-210 was observed for ~77 days at a ≈30 minute cadence by the *Kepler* space telescope during Campaign 2 of the *K2* mission (Howell et al. 2014). Systematic artifacts in the photometry due to spacecraft attitude adjustments were corrected for using an established algorithm (Aigrain et al. 2016) and the resulting time series is shown as the top sequence in Figure 2. The deep dimming events are superposed on the smooth rotation signature of the spotted star, apparently in phase with the stellar rotation. Figure 3 highlights the dimming events in the context of rotation pattern.

Our analysis and interpretation of the dimming events is predicated on the following observations and assumptions, all of which we develop in full below:

1. Dimming events of variable depth, duration, and morphology occur every 5.6685 days, in phase with the stellar rotation.

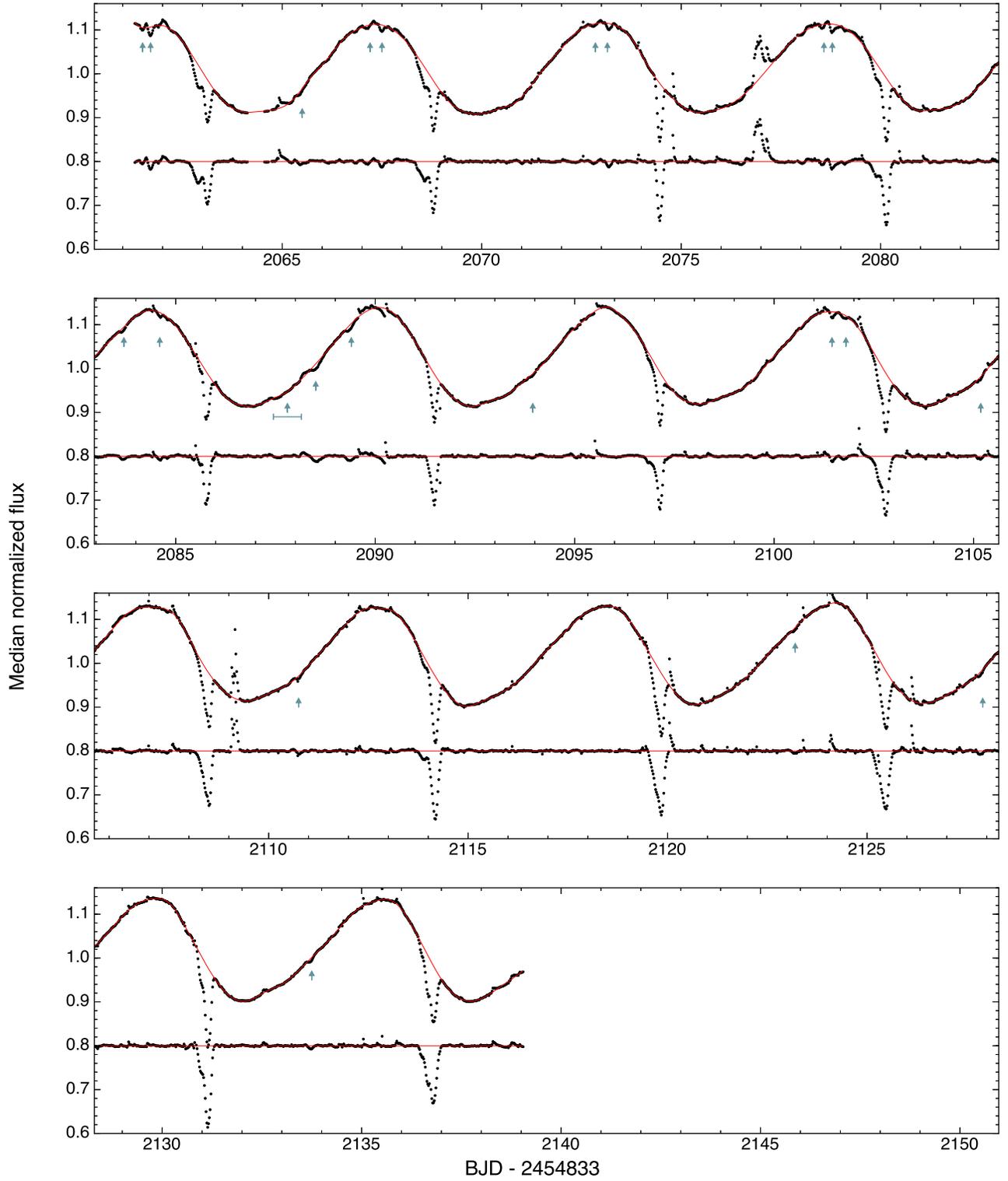

**Figure 2.** *K2* light curve of RIK-210 in 20-day segments (black points). The red curve indicates the iterative spline fit (variability fit A) used to remove the starspot modulation pattern. Blue arrows indicate the approximate positions of shallow dimming events. Below the light curve we plot the residuals of the variability fit, normalized to unity but shifted to a continuum value of 0.8 in this figure. A broad flux dip occurs near BJD - 2454833 = 2087.8, indicated by the arrow and bar in the second panel, however the variability fit shown here passes through it and so it is not apparent in the residuals.

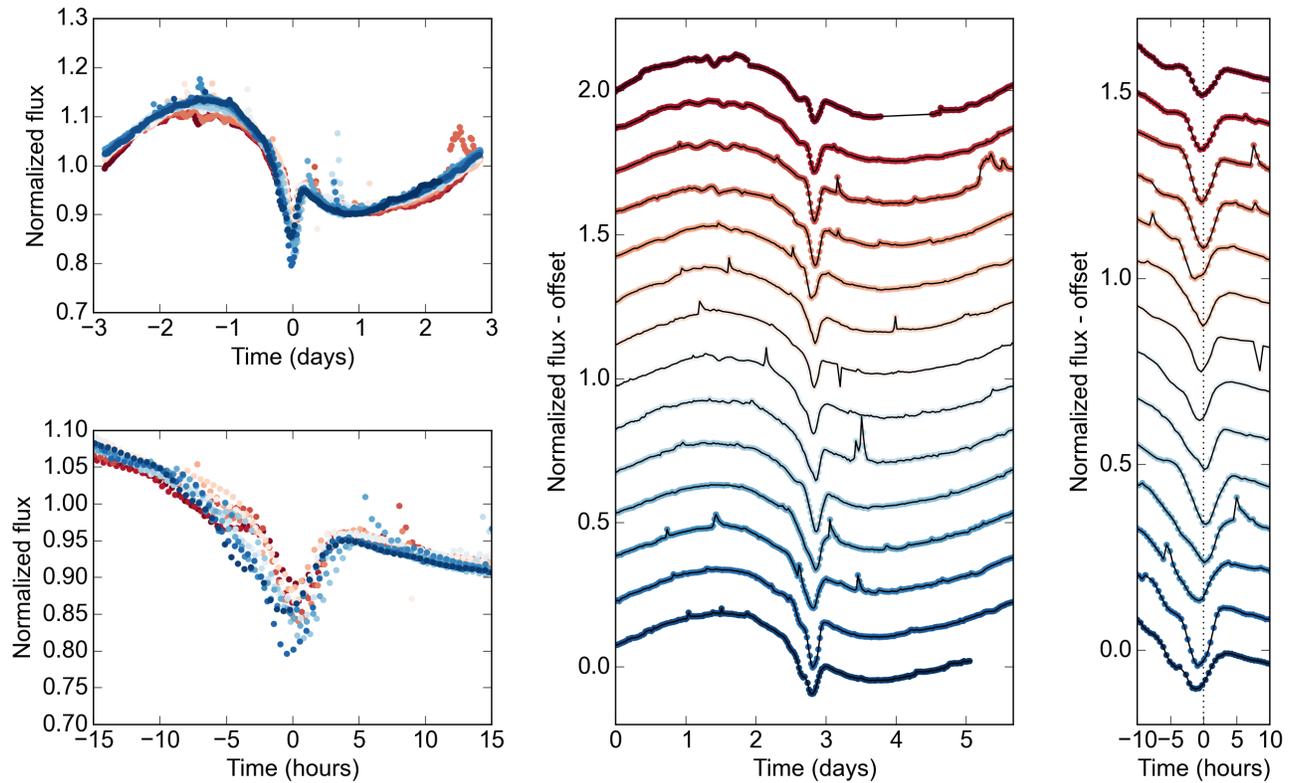

**Figure 3.** *K2* light curve of RIK-210. In each panel, the point color indicates the relative time of observation (with red corresponding to earlier times). Upper left: photometry folded on the rotational/orbital period of 5.67 d. Bottom left: same as above, showing an enhanced view of the dimming events. Middle: Same as the upper left, but with vertical offsets applied after each rotation of the star. Right: Same as middle, showing an enhanced view of the dimming events. There is clear evolution in the depth, duration, and overall morphology of the dimming events, a strong indication against a transit by or eclipse of a single solid body of any size. Some flares appear to occur at approximately the same rotational phase, shortly after the transit events.

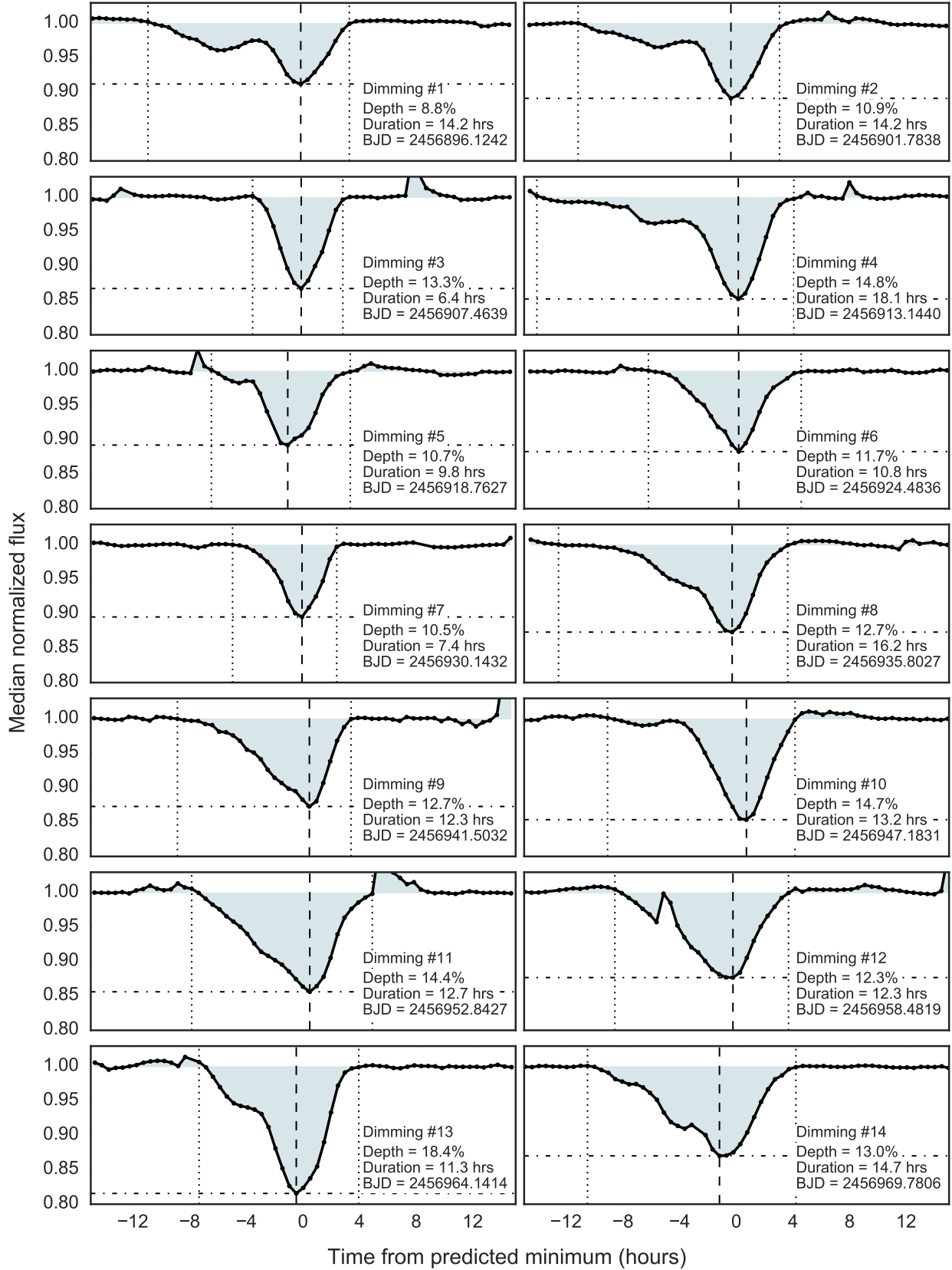

**Figure 4.** Individual dimming events within the *K2* light curve of RIK-210, after removing the starspot modulation. On the abscissa, the data are plotted in terms of time from the predicted minimum as determined from a linear ephemeris. The maximum depth, total duration, and time of minimum light are summarized in the lower right corner of each panel.

2. The dimming events are both deep (sometimes greater than 15%) and short in duration relative to the rotational period, and thus unlikely to be due to features on the stellar surface.

3. The morphology of the dimming events is variable over the 77 day campaign, while the starspot modulation pattern remains stable over this timeframe.

4. If the dimming events are due to an object or debris cloud in a Keplerian orbit around the star, the transiting body must be at or near the corotation radius.

5. Assuming the transiting material does not contribute significant flux to the optical light curve, the depth of each transit yields the approximate size of the occulting body or bodies.

6. The facts that the dimming events are periodic yet narrow in rotational phase indicates the occulting material is azimuthally confined within its orbit around the star, else if it is arranged in a torus-like structure its orbital plane is tilted and its distribution vertically above that plane is inhomogeneous.

7. A purely gaseous cloud lacks the requisite opacity in the *Kepler* bandpass to produce the observed transit depths.

### 3.1. *Stellar variability fit*

At several stages of our analysis, we consider a "flattened" light curve, in which we attempted to remove the underlying starspot modulations while excluding dimming events from the fit. We use two fits in this work. The first, which for convenience we call variability fit A, was achieved using a custom, automated routine which iteratively fits the light curve via a cubic basis spline, excluding $2\sigma$ outliers upon each iteration (for details see David et al. 2016a). The second, which we call variability fit B, also utilizes a cubic basis spline, though the observations excluded from this fit were manually selected through visual inspection in a cadence-by-cadence manner. We note that fit A is used for quantitative analyses and fit B for qualitative or illustrative purposes. The adopted fit and the residual light curve are displayed as the middle and bottom sequences in Figure 2.

### 3.2. *Periodicity*

A period of $5.670 \pm 0.004$ d that we associate with the rotation of the star was determined from a Lomb-Scargle periodogram of the systematics-corrected light curve. We note this rotation period is somewhat long for Upper Sco members of comparable color but not outside the distribution, which peaks at periods shorter than 1 d with a long tail out to tens of days (L. Rebull, priv. comm.). The rotation period uncertainty was estimated as $\Delta P = \sigma\sqrt{2/M}$, where $\sigma = dt/2$ is half of the observing cadence and $M$ is the number of rotational cycles contained within the entire light curve (Mighell & Plavchan 2013). Performing the same periodogram analysis on only the in-transit observations, or on only the out-of-transit observations, yields identical periods within uncertainties. We conclude that the orbital period is indistinguishable from the rotation period of the star, indicating the occulting material is orbiting at or near the corotation radius.

The corotation radius is the separation at which an object in circular orbit around a star has an orbital period equivalent to the stellar rotation period, given by:

$$a_{\rm corot} = (M_*/M_\odot)^{1/3}(P_{\rm rot}/{\rm yr})^{2/3} \text{ AU.} \quad (1)$$

For stars still hosting massive protoplanetary disks, theory predicts truncation of the inner disk near the corotation radius, where ionized material is dragged along magnetospheric field lines and may be accreted onto the stellar surface (König 1991; Collier Cameron & Campbell 1993; Bouvier et al. 1997). For the adopted stellar mass and the rotation period above, we find $a_{\rm corot} = 0.050 \pm 0.004$ AU, or approximately $9~R_*$.

### 3.3. *Ephemeris*

From the *K2* light curve, we determined an ephemeris for the dips from $10^6$ bootstrapping simulations of linear fits to the times of minimum light, conservatively assuming 0.5 hr uncertainties and Gaussian errors. The times of minimum light are predicted by:

$$t_n = 2456896.1278(0.011) + n \times 5.6685(0.0014) \text{ BJD.} \quad (2)$$

Notably, the time of minimum light deviates from this linear ephemeris within the *K2* campaign, sometimes occurring earlier or later by up to one hour. This is potentially due to morphological changes of the obscuring body itself rather than dynamical effects, as discussed in § 3.8.

### 3.4. *Durations*

Figure 4 highlights the individual dimming events during the *K2* campaign. Assuming an edge-on and equatorial transit, the transit duration of a massless, dimensionless particle orbiting with circular velocity $v_{\rm circ}$ and period $P_{\rm orb}$ at a distance $a$ from a star with mass and radius, $M_*$ and $R_*$, respectively, is given by:

$$t_{\rm cross} = \frac{2R_*}{v_{\rm circ}} = \frac{2R_*}{(GM_*/a)^{1/2}} = \frac{P_{\rm orb}}{\pi}\frac{R_*}{a}. \quad (3)$$

For a particle on a circular orbit at the corotation radius, the transit duration is:

$$t_{\rm cross} = \frac{2R_*}{G^{1/2}(M_*/P_{\rm orb})^{1/3}}. \quad (4)$$

For the stellar mass and radius adopted above, and the presumed orbital period, the expected transit duration is $\approx 5$ hours, i.e. slightly less than the minimum observed duration but about three times shorter than the maximum observed duration. A larger transiting object would produce longer durations, though the variable transit duration, depth, and morphology is inconsistent with eclipses by a single spherical body. We conclude that the occulting material must be comprised of, at least in part, an extended distribution of particles with small individual size, but large collective size, relative to the star.

### 3.5. Size of occulting material

A spherical body of radius $r$ transiting the equator of a star with radius $R_*$ has a transverse velocity of

$$v_t = \frac{2(r + R_*)}{t_{\mathrm{dim}}}, \tag{5}$$

where $t_{\mathrm{dim}}$ is the duration of a transit or eclipse.

If the object is on a circular orbit, the transverse velocity is the circular Keplerian velocity, and the size of the occulting object is given by,

$$r = \frac{t_{\mathrm{dim}}}{2}\left(\frac{GM_*}{a}\right)^{1/2} - R_*. \tag{6}$$

From Kepler's third law, assuming the object in orbit has a mass much smaller than the mass of the star, the radius of the occulting body is then,

$$r = \frac{\pi a t_{\mathrm{dim}}}{P_{\mathrm{orb}}} - R_*. \tag{7}$$

For the range of dimming durations observed in RIK-210 ($\approx$6–18 hours), the radius of a putative occulting body would be in the range of $\sim$0.3–3 $R_\odot$ or $\sim$3–30 $R_{\mathrm{Jup}}$, assuming an equatorial orbit. However, we note that the dips with the longest observed durations are always significantly asymmetric. A single, spherical body produces a symmetric transit or eclipse profile (in the absence of spot crossings or precession). Thus, the larger end of the radii quoted above likely represent the size of a putative debris stream or dust cloud rather than any individual occulter.

If instead we assume the occulting material is distributed azimuthally in orbit at the corotation radius around the star, we can calculate its linear size from its arclength as implied by the dimming durations:

$$s = \frac{2\pi a_{\mathrm{cor}} t_{\mathrm{dim}}}{P_{\mathrm{orb}}}, \tag{8}$$

from which we found $s \sim$3–9 $R_\odot$, i.e. larger than the diameter of the star itself. This would imply the dust or debris covers $\sim$4–13% of its orbit, or an angle of 0.3–0.8 radian, assuming equatorial transits.

### 3.6. Dip morphology

The dip morphology continually evolves throughout the K2 campaign (Figure 3 and Figure 4). While no two dips have the same morphology, there are common characteristics between all events, which are evident in Figure 5. Notably, ingress is often shallower than egress. Such dip morphologies may be produced by a leading tail of debris surrounding a main transiting body, or by a precessing body transiting a star of non-uniform surface brightness. Additionally, the variance in the phase folded light curve shows a double-peaked profile, gradually increasing through ingress before reaching a local minimum at the expected time of mid-transit, a global maximum shortly after the time of mid-transit, and then falling off steeply in egress.

In some individual dips, there are clearly two or even three distinct minima, with the shallower minima always occuring prior to the global minimum (Figure 4). Additionally, some dips exhibit either ingress or egress morphologies that are strikingly similar to the morphologies expected from transiting comets (Lecavelier Des Etangs et al. 1999; Lecavelier Des Etangs 1999).

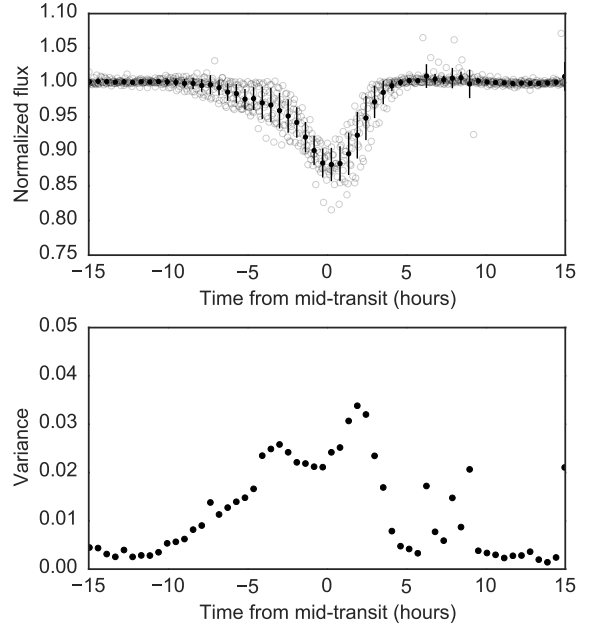

**Figure 5.** Top: Phase folded and binned K2 light curve of RIK-210 after removing the starspot rotation signal via an iterative spline fit. Bottom: variance in the phased and binned light curve. Variability fit B was used to make this plot.

### 3.7. Bursts, Flares, and Starspots

The out-of-transit light curve is dominated by a semi-sinusoidal waveform of $\sim$20% peak-to-trough amplitude, which we interpret as rotational modulation of starspots. While the starspot modulation amplitude is large even for a young star, it is not unprecedented. For example, a spot modulation amplitude of $\Delta V \approx 0.8$ mag has been observed in the weak-lined T Tauri star (WTTS) LkCa 4 (Grankin et al. 2008). The brightness difference observed in RIK-210 suggests a $\sim$5% difference in the disk-averaged temperature (i.e. a few hundred Kelvin) between the coolest and hottest hemispheres.

In addition to the flux decrements, there are several flares in the light curve, which are commonly observed in young, low-mass stars. There is marginal evidence that these flares are somewhat confined in rotational phase (see top left panel of Figure 3). There is also a large burst, approximately 12 hours in duration, occurring at BJD = 2456909.78 (or BJD - 2454833 = 2076.78, as shown in Fig. 2 for example). This burst is different in morphology from the typical flare signature, and could be due to discrete accretion onto the star, possibly a low level form of the events illustrated in Cody et al. (2017) for actively accreting stars observed in the same K2 campaign. However, as noted above, the star lacks both spectroscopic accretion indicators and an infrared excess indicative of an inner disk. Notably, the end of the burst feature is characterized by two consecutive flare-like decays.

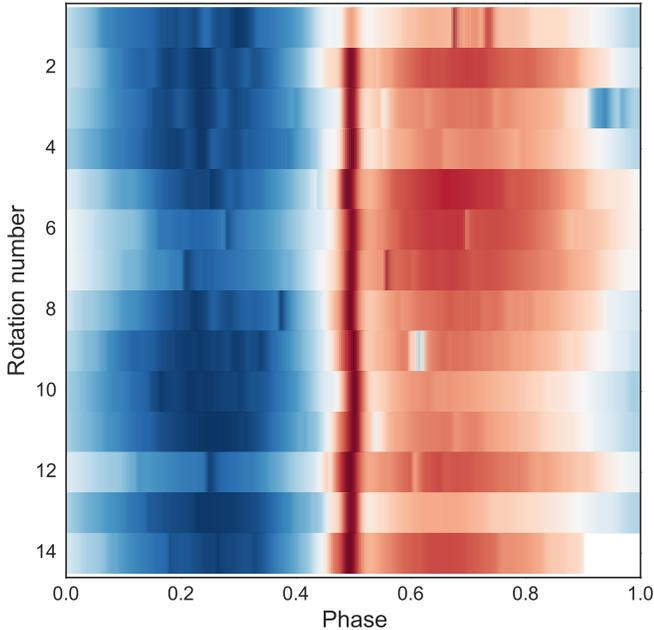

**Figure 6.** Waterfall diagram of RIK-210. Colors represent the *K2* light curve intensity, with blue corresponding to higher flux and red to lower flux. The primary dimming events that are the focus of this work are clearly seen as the dark red stripe, which changes in intensity, duration, and timing. At top left, a series of light blue stripes between phases of 0.2 and 0.3 are observed to apparently drift in phase over the first four cycles. These are some of the shallow flux dips discussed in § 3.9. Linear interpolation was performed over data gaps where spacecraft thruster firings were excluded.

### 3.8. *Light curve parameterization*

We characterize the *K2* dimming events in terms of eight parameters: maximum depth, total duration, area (the time-integrated extinction), time variation between predicted and observed minimum light, the slopes of ingress and egress, and the durations of ingress and egress (see Table 2).

All parameters were determined from the flattened light curve, using variability fit A (Figure 2). No interpolation was performed within individual transits, and as such the maximum depth and time of minimum light were simply determined by the observation with the lowest flux. The total duration was determined by finding the last cadence before mid-transit, and the first cadence after mid-transit, to fluctuate above a median normalized flux of unity. The slope of ingress and egress are determined from the 6 hours preceding, and 3 hours proceeding the time of minimum light, respectively. Often the ingress clearly has multiple minima, which informed the decision to measure the slope for only a portion of ingress.

There is no observed correlation between the depths and durations of dimming events, as was observed for the dips ascribed to a transiting planet around the young star PTFO 8-8695 (Yu et al. 2015). For a single spherical body transiting the disk of a star, both the transit depth and duration depend on the size of the occulting body. Alternatively, if dust is partially contributing to the dimming events in RIK-210, the depths may change independently of the durations due to changing optical depths.

However, there is a correlation between the slope of egress and the timing variation, such that the later the time of minimum light occurs, the steeper the egress (Figure 7). The Spearman rank correlation coefficient is $r = 0.617$, with $p < 2\%$. The significance of this correlation increases ($r = 0.65$, $p < 1.2\%$) if variability fit B is used to flatten the light curve. From $10^5$ bootstrapping simulations we estimate the correlation to be significant at the $\approx 2.2\sigma$ level, and estimate the true correlation coefficient $r$ to be between 0.4–0.7 at 95% confidence assuming 0.25 hr uncertainties in the timing variations. We note that the correlation strengthens and becomes more significant depending on how the egress slope is measured, and the above estimates are somewhat conservative.

To investigate the significance of the timing variations, we measured the "transit" times using two methods: 1) Gaussian fits to the deepest component of each dip, considering data within 2.4 hours of the faintest cadence, using the Levenberg-Marquardt algorithm, and 2) selecting the cadence with the lowest flux value in each dip. We found that the timing variations with respect to a linear ephemeris agree (in both magnitude and direction) in most cases to within 0.25 hours, regardless of the technique used to determine the individual "transit" times.

Notably, there is no correlation between the slope of ingress and the timing variation, nor is there one between the egress duration and the timing variation. Timing variations are most typically the result of gravitational interactions between multiple orbiting bodies. However, given the asymmetries of the dimming events in RIK-210, it is also possible that the observing timing variations are due to the occulting body or bodies changing shape, e.g. variable extinction by a dust cloud. Such a scenario could displace the time of minimum light and mimic timing variations from gravitational interactions.

One possible explanation for the timing variation-egress slope correlation involves a collection of bodies which are gravitationally interacting. A massive body in a trailing orbit with respect to a less massive collection of bodies might pull the leading bodies back in their orbit, causing the configuration to become more spatially compact and delaying the transit time. The more compact the configuration, the less dispersion one expects in stellar disk-crossing times between component bodies and the steeper egress will be. An issue with this explanation is that one might also expect steeper ingress. However, given that some component of the occulting material may be dust, it is not evident whether the timing variations are tracing dynamical evolution or morphological evolution (i.e. due to a variable dust cloud).

It is interesting to note that depth (and apparently morphology) variations which correlate with stellar rotational phase have been observed in the transits of the disintegrating planet KIC 12557548b (Kawahara et al. 2013; Croll et al. 2015). Additionally, Croll et al. (2015) found morphological variations that correlated with the transit timing variations, such that late transits had shallower dips with a more gradual egress relative to the early or on-time transits. Both studies investigated the possibilities that the correlation between transit depth and rotational phase is due to (1) spot-crossing events by the cometary tail (which leads to anomalous brightening if the tail occults a starspot, thereby changing the transit morphology and inducing an apparent timing variation), or (2) periodically modulated mass-loss due to an active region on the star with enhanced X-ray and UV flux.

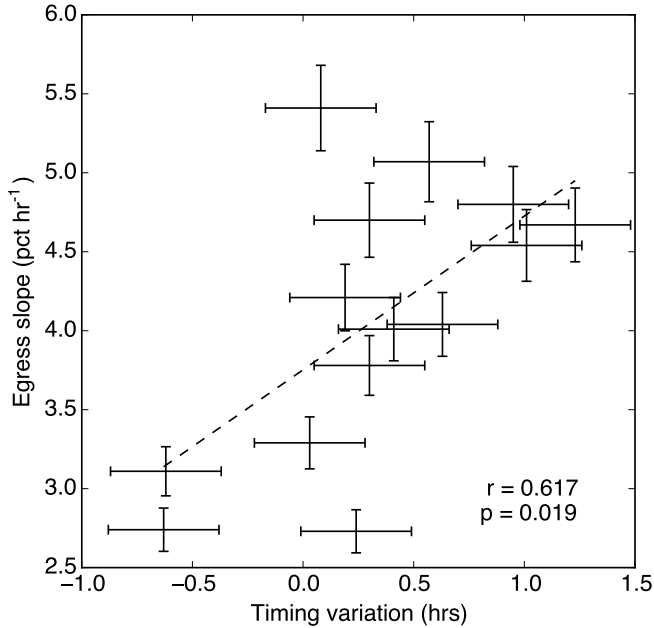

**Figure 7.** Correlation between the dimming timing variatons and egress slopes. *Ad hoc* uncertainties of 0.25 hr in timing variation and fractional uncertainties of 10% in slope are assumed.

For RIK-210, the obscuring material, if in orbit about the star, has the same period as the stellar rotation. Thus, while RIK-210 is heavily spotted, spot-crossing anomalies are not a satisfactory answer for the transit variations, unless individual spots are rapidly evolving or different regions of the stellar disk are being occulted from one transit to the next (perhaps as a result of an an obscuring cloud that is evolving in shape or crossing different stellar latitudes as it transits the disk of the star).

To qualitatively assess the similarity of individual dimming events to solid-body transits, we performed Levenberg-Marquardt fits to the observations using the JKTEBOP light curve modeling code (Southworth 2012, and references therein). All fits presented here assumed a linear limb-darkening coefficient of 0.6 for the primary star, allowing the following parameters to vary: time of mid-transit, inclination, ratio of radii, and the sum of the radii divided by the semi-major axis. These fits typically favor low inclinations ($\sim 81°$), a ratio of radii of $\approx 0.48$, and a sum of fractional radii of $\approx 0.22$. However, these parameters are degenerate and the resulting values are not given significant weight in our interpretation of the data. We also note the size ratio and sum of fractional radii imply conflicting sizes of the occulter. Nevertheless, these models are useful for assessing the symmetry and underlying geometry of the occulting material.

The first dimming event is also the most shallow. Interestingly, by fitting the deepest component of the first dimming event (Figure 8), and subsequently fitting this model to the rest of the *K2* data (allowing only the ephemeris timebase, $T_0$, and period to vary), we find that essentially all other observations fall *below* this model (Figure 9). One might interpret this observation as evidence for a primary occulting body surrounded by an evolving swarm of dust or debris. The residuals of this "minimum-obscuration fit" are quite often double-peaked (Figure 10), perhaps suggestive of multiple obscuring bodies or a single, clumpy occulter.

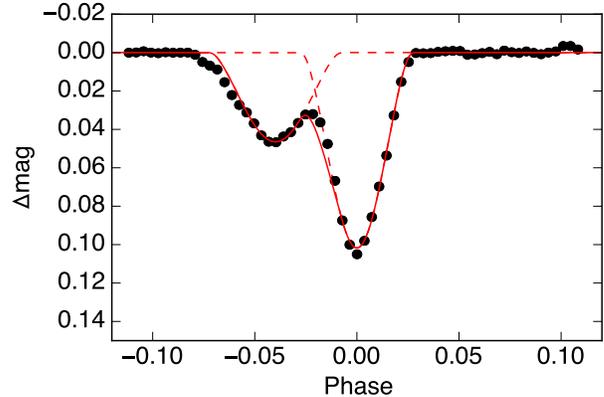

**Figure 8.** Composite of two independent JKTEBOP model fits to the two minima in the first dimming event of RIK-210 in the *K2* photometry. At least some of the individual transits observed by *K2* are reasonably well fit by one to three spherical occulting bodies with large size relative to the star. The fit to the deeper dip alone is referred to later as the "minimum obscuration fit." Variability fit B was used to make this figure.

### 3.9. *Short duration, shallow flux dips*

In addition to the main dimming events of several to $\sim 20$ percent depth, shallower transit-like dips are also apparent in the *K2* light curve (Figures 2 and 6). In general, these dips are prominent in the beginning of the time series (when the primary dips are generally shallower), and apparently absent by the end of the campaign. These dimming events are typically $\sim 5$ hrs in duration, consistent with the expected transit duration for an orbiting body at the corotation radius.

In the beginning of the *K2* campaign, there is a grouping of at least four consecutive dips, which persists to some extent for the first four orbits/rotations (Figures 2 & 12). Over the next four orbits, the spacing between prominent dips appears to increase, somewhat suggestive of bodies that have departed from the original periodicity of $\approx 5.67$ days. Finally, over the remainder of the campaign, the number of shallow dips appears to gradually decline. There is at least one dip present in each of the last six orbits/rotations. However, because this dip is not strictly periodic, it is not clear whether this feature is attributed to the same body, or multiple bodies at different semi-major axes. If the former case is true, then the time variations between transits ($\sim 0.6$ d) would represent a fractional transit timing variation of unprecedented amplitude.

In at least one case, a transit of $\approx 0.7\%$ depth appears to recur with a period similar, but not equal to, the deep dimming events (Figure 11). Some of the shallow transits appear to have a characteristic *trailing* tail morphology, as opposed to the deep dimming events which possess a morphology reminiscent of a *leading* tail. As Rappaport et al. (2012) points out, in cases such as these the transit depth does not simply reflect the underlying planet size

**Table 2**
Parameters of the transient transit signature

| Event | $T_{min}$ BJD | Depth (%) | Duration (hrs) | Area (% hrs) | Timing variation (hrs) | Ingress slope (%/hr) | Egress slope (%/hr) | Ingress duration (hrs) | Egress duration (hrs) |
|---|---|---|---|---|---|---|---|---|---|
| 1 | 2456896.1242 | 8.80 | 14.22 | 50.76 | 0.24 | -0.85 | 2.73 | 10.79 | 3.43 |
| 2 | 2456901.7838 | 10.89 | 14.22 | 56.16 | 0.03 | -1.22 | 3.29 | 10.79 | 3.43 |
| 3 | 2456907.4639 | 13.26 | 6.37 | 42.47 | 0.30 | -2.38 | 4.70 | 3.43 | 2.94 |
| 4 | 2456913.1440 | 14.78 | 18.14 | 76.21 | 0.57 | -2.07 | 5.07 | 14.22 | 3.92 |
| 5 | 2456918.7627 | 10.74 | 9.81 | 41.99 | -0.62 | -1.87 | 3.11 | 5.39 | 4.41 |
| 6 | 2456924.4836 | 11.69 | 10.79 | 47.26 | 0.63 | -2.08 | 4.04 | 6.37 | 4.41 |
| 7 | 2456930.1432 | 10.46 | 7.36 | 32.63 | 0.41 | -1.91 | 4.01 | 4.90 | 2.45 |
| 8 | 2456935.8027 | 12.66 | 16.18 | 66.04 | 0.19 | -1.72 | 4.21 | 12.26 | 3.92 |
| 9 | 2456941.5032 | 12.73 | 12.26 | 63.32 | 0.95 | -1.90 | 4.80 | 9.32 | 2.94 |
| 10 | 2456947.1831 | 14.65 | 13.24 | 63.33 | 1.23 | -2.86 | 4.67 | 9.81 | 3.43 |
| 11 | 2456952.8427 | 14.40 | 12.75 | 83.37 | 1.01 | -1.88 | 4.54 | 8.34 | 4.41 |
| 12 | 2456958.4819 | 12.33 | 12.26 | 67.52 | 0.30 | -2.06 | 3.78 | 8.34 | 3.92 |
| 13 | 2456964.1414 | 18.44 | 11.28 | 88.17 | 0.08 | -2.72 | 5.41 | 6.87 | 4.41 |
| 14 | 2456969.7806 | 12.97 | 14.71 | 84.57 | -0.63 | -1.72 | 2.74 | 9.32 | 5.39 |

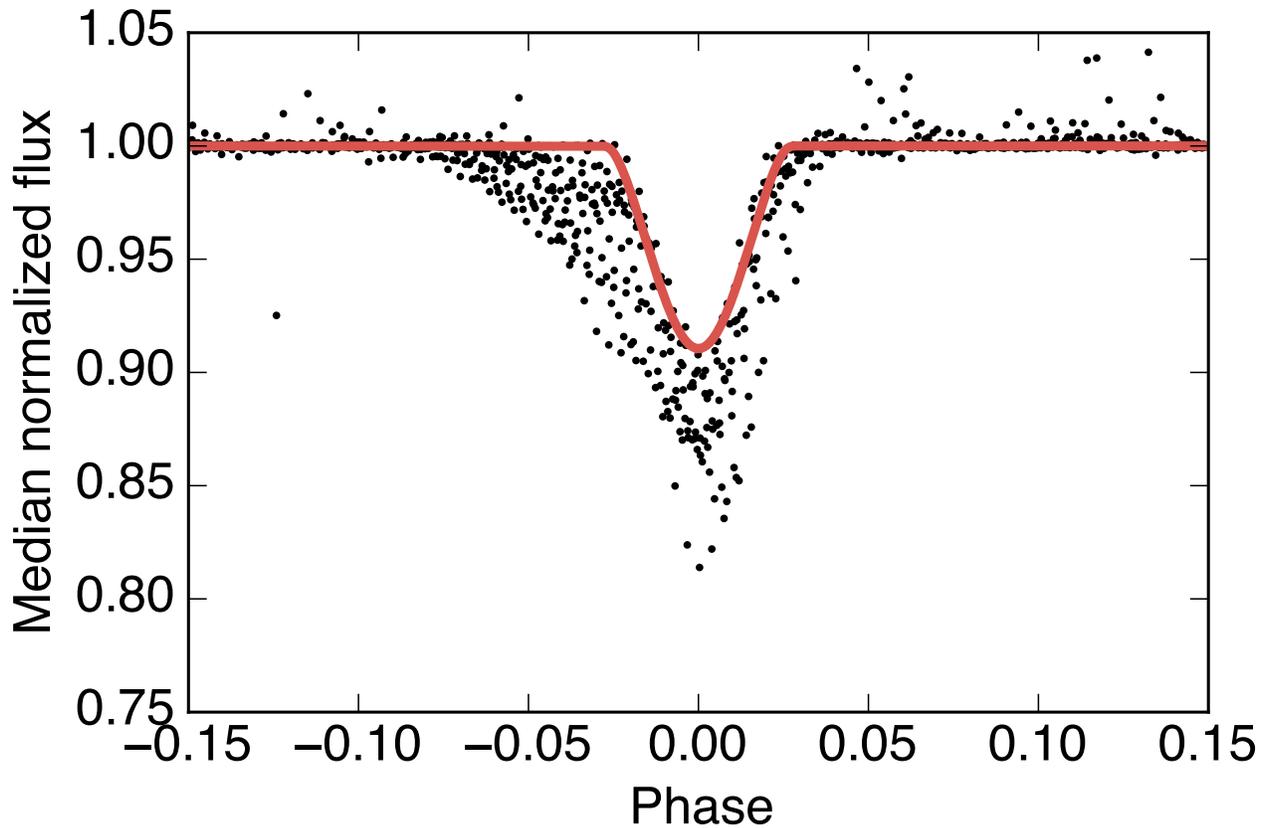

**Figure 9.** JKTEBOP model (red) overplotted on the flattened *K2* light curve of RIK-210 (black points). The model above was fit to the deepest component of the first dimming event in the *K2* campaign, then refit to the entire light curve allowing only the period and time of mid-transit to vary. Few observations lie above this fit, perhaps suggestive of an underlying, spherical occulting body surrounded by a stream of dust or swarm of planetesimals. Variability fit B was used to make this figure, though the conclusions drawn are independent of the fit used.

but also the size of a putative cloud of dust. Thus, the transit depth may be expressed as,

$$\delta \sim \left(\frac{f R_{\rm Hill}}{R_*}\right)^2 \approx \left(\frac{f (M_P/3M_*)^{1/3} a}{R_*}\right)^2, \quad (9)$$

where $R_{\rm Hill}$ is the planetary Hill radius, $f$ is the fraction of the Hill radius that is optically thick, and $M_P$ is the planet mass. Inverting the above equation, we can crudely approximate the planet masses required to create a 1% transit. Assuming $a$=0.05 AU, the planet masses required are: 2.6 $M_{\rm Jup}$ for $f = 0.1$, 6.5 $M_\oplus$ for $f = 0.5$, or 1.6 $M_\oplus$ for $f = 0.8$.

It is worth noting that the initial grouping of four or more dips are located near the light curve maximum and approximately 90 degrees out of phase with respect to the primary dimming events. By comparison, Trojans are concentrated around the L4 and L5 Lagrange points, 60 degrees ahead and behind the primary orbiting body, respectively. Trojan populations do have an azimuthal distribution, and some of Jupiter's Trojans do in fact orbit 90 degrees ahead or behind the giant planet. However, we also note that a stable Trojan population would have the same period as the main occulter, and this does not seem to be the case for the shallow dips which do not exhibit strict periodicity.

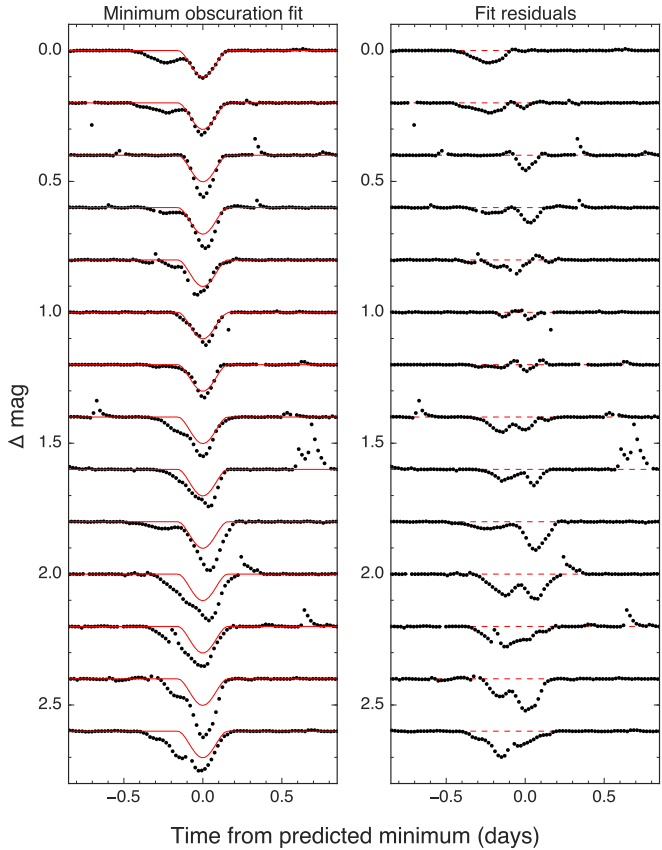

**Figure 10.** Evolution of the primary dips in the *K2* light curve of RIK-210. In both panels, the top row shows data from the first dip, proceeding consecutively downwards to the bottom row, which shows the last observed dip. *Left:* the "minimum obscuration fit," obtained by fitting a model to the deepest component of the first dimming event, with respect to subsequent individual transits. *Right:* evolution of the residuals to the fits depicted on the left.

In any event, the fact that these shallow dips are not strictly periodic suggests that they are not due to features on the surface of the star nor can they be due to material that is strictly co-rotating with the star or stellar magnetosphere. We posit that these shallow dips are likely related to the deeper dimming events, and speculate that they are due to transiting material in orbit about and nearly corotating with the star.

It is revealing to compare the behavior of these shallow dips to the dips observed around the white dwarf WD 1145+017 (Vanderburg et al. 2015). The current interpretation for that system is of a large asteroid transiting a white dwarf and orbiting near its Roche limit. As a result of tidal disruption, fragments break off from the asteroid, producing distinct flux dips in time series photometry which have been noted to "drift" in phase relative to the transits of the parent body (Rappaport et al. 2016).

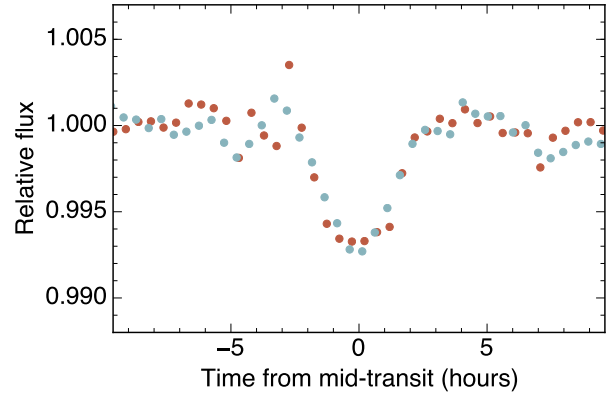

**Figure 11.** Example of shallow dimming events seen late in the *K2* campaign. Here we show two events (indicated by different colored points) separated by 5.888 days with consistent depth, duration, and morphology.

## 4. ARCHIVAL AND FOLLOW UP PHOTOMETRIC MONITORING

Given the variability in the depth, duration, morphology, and timing of the dimming events observed by K2, we sought auxiliary time series photometry to investigate changes over longer time baselines. There is no archival light curve for RIK-210 from HATSouth as that survey avoids crowded fields (G. Bakos, priv. comm.). Fortunately, there are more than 28,000 individual observations between 2006–2010 from SuperWASP-South (Figure 13), hereafter WASP (Pollacco et al. 2006).

A coherent Fourier transform of all the WASP data yields a high-significance peak with only a small first harmonic, indicative of sinusoidal behavior. The period favored by the WASP data is $P = 5.5665 \pm 0.0012$ days, within $1.6\sigma$ of the period determined from the *K2* data (where $\sigma$ here represents the uncertainties of the two periods summed in quadrature). Thus, WASP clearly detected the rotation period of the star, but there is mixed evidence regarding the presence of deep flux dips in prior years. A notable feature of the combined WASP light curve, presented in the left panel of Fig. 13, is a broad depression at phases earlier than the expected dip phase from *K2*, seemingly suggestive that the dip was present in previous years but drifted in phase, narrowed, or both. Upon first glance of the annual WASP light curves (right panel of Fig. 13), there appear to be some statistically significant dips in previous years (e.g. near phase of 0.65 in 2008). However, further inspection of the WASP data reveals that most of the dip-like structures seen in prior years are the result of data acquired on only one or two nights in that year. The most complete phase coverage was achieved in both 2009 and 2010, but prominent dips are not readily apparent. We conclude that there is no strong evidence for dips in the 2006–2010 WASP data, and suggest the broad trough in the combined light curve may be the result of a spurious dip in 2006 and/or evolution of the spot pattern.

We also acquired follow up ground-based photometry of RIK-210 in the Bessel *V* and Sloan *i'* filters using the Las Cumbres Observatory Global Telescope network, hereafter LCOGT (Brown et al. 2013), between UT 2016-07-06 and 2016-08-10. Raw images were automatically processed using the LCOGT BANZAI pipeline, which performs bad pixel masking, bias and dark subtraction, flat field correction, and provides an astrometric solution. We performed differential photometry from the reduced images with AstroImageJ (Collins et al. 2016), using 2MASS J16234525-1722086 and 2MASS J16230556-1716209 as comparison stars. The color of RIK-210 in these bands is $V - i'$=1.77±0.17 mag (uncorrected for reddening, from APASS DR9). This is a bluer color than expected given the M2.5 spectral type, and corresponds more closely to a K9 type photosphere. However, this color discrepancy is not apparent from the SED in Figure 1.

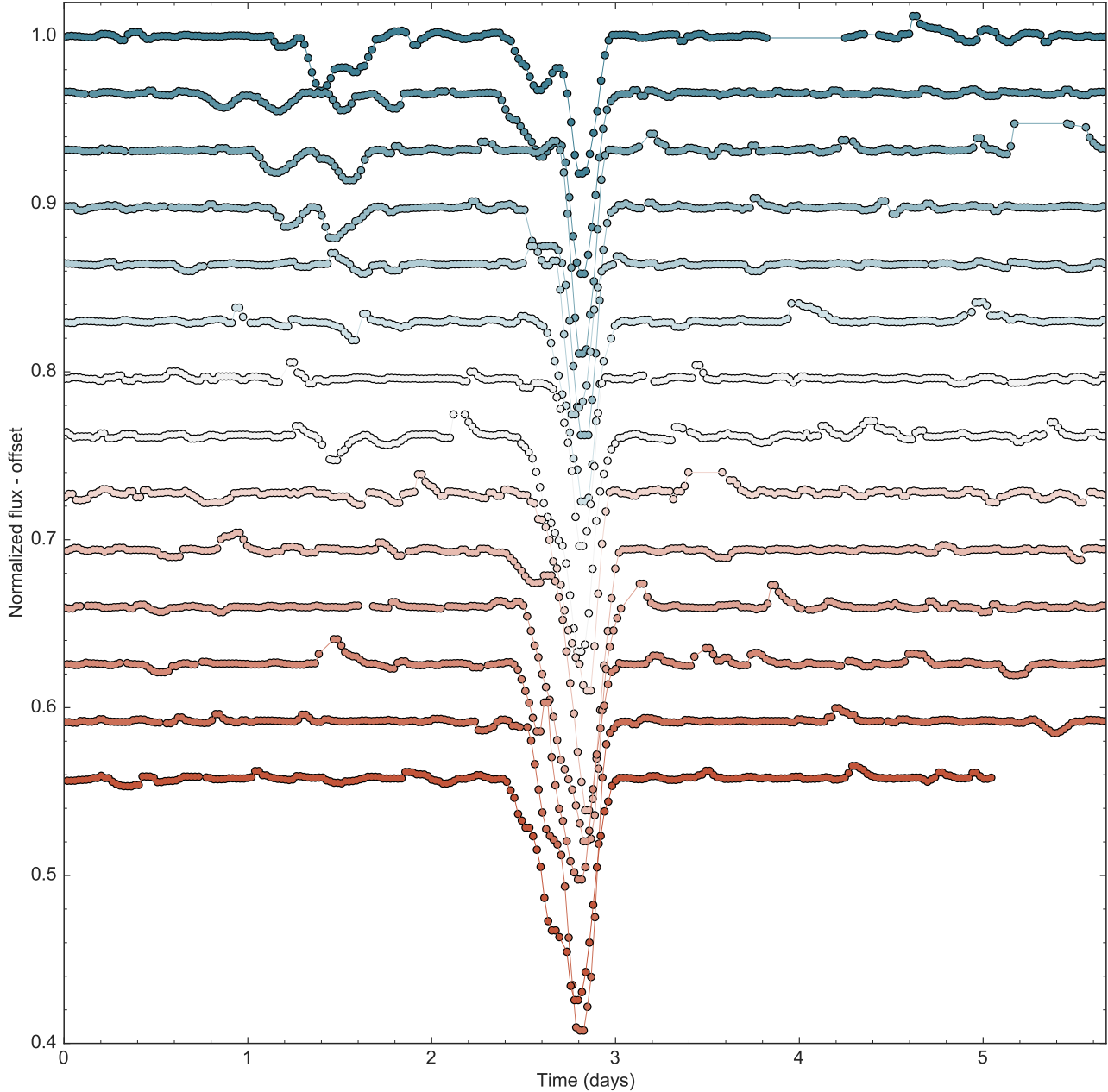

**Figure 12.** Flattened and median filtered *K2* light curve of RIK-210. The data are phase-wrapped on the dip period of 5.6685 days. A grouping of shallow dips preceding the main dip is prominent for the first four rotation periods of the campaign, then largely disappears. Variability fit B was used to make this figure.

Conspicuously, observations from LCOGT at the predicted time of the dimming events (phase=0.5 in Figure 13) show that no such dimming events are observed, at least not at the ∼0.2 mag level observed by *K2*. The absence of a transit or eclipse at the time predicted from a linear ephemeris based on the *K2* data is in tension with the explanation that the dimming events could be due to a solid body with any substantial size relative to the star. We note that because phase coverage is sparse around the predicted time of the dimming events, and due to the precision of our ground-based photometry, we can not rule out the possibility that dimming events of a few percent depth are still occurring.

The LCOGT photometry (Figure 13) confirmed the ephemeris of the stellar rotation has remained unchanged more than 2 years after the K2 observations, though the amplitude of variability in both the *V* and *i'* bands is notably higher than that observed by K2 (the Kepler bandpass corresponds roughly to the *V* and *R* filters). Con-sidering the WASP, *K2*, and LCOGT data together (Figure 13), the amplitude of variability due to spot modulation appears to have increased monotonically from 2006 to 2016.

The $V - i'$ color varies sinusoidally as a function of rotational phase, with the star becoming bluer when brighter and redder when fainter, as one expects from starspots. The color as a function of rotational phase is used to estimate the temperature contrast between the coolest and hottest hemispheres of the stars. The peak to trough amplitude is ≈0.2 mag, corresponding roughly to the $V - i$ color difference between a single spectral type class for M-type pre-main sequence stars (Pecaut & Mamajek 2013). This corresponds to a temperature difference of ∼100–200 K, which is consistent with spot temperature contrasts published for M-dwarfs of similar temperature (Andersen & Korhonen 2015). The temperature ratio between the coolest and warmest hemispheres can also be estimated from the peak-to-trough amplitude of the *K2* light curve and the Stefan-Boltzmann law:

$$\frac{L_{\min}}{L_{\max}} = \left(\frac{T_{\min}}{T_{\max}}\right)^{1/4} \sim 0.8, \qquad (10)$$

consistent with the temperature difference estimated from the $V - i'$ variability amplitude. We note for a given amplitude of photometric modulation, the spot temperature and size are degenerate parameters, but that the above considerations indicate the difference between the average hemisphere temperatures.

## 5. SPECTROSCOPIC OBSERVATIONS

We acquired multiple high-dispersion spectra of RIK-210 with Keck-I/HIRES (Vogt et al. 1994), covering either ∼3600-8000 Å or 4800-9200 Å depending on the spectrograph settings. These spectra were used to perform a search for secondary lines due to a companion, measure the projected rotational velocity, and monitor variations in the radial velocity and line profiles of RIK-210.

### 5.1. *Secondary line search*

A search for secondary lines in the spectrum of RIK-210 was performed following the procedure of Kolbl et al. (2015). The method involves fitting template spectra of >600 FGKM stars (mostly dwarfs or subgiants) observed with Keck/HIRES to the continuum-normalized spectrum of RIK-210, then fitting the best-fit residuals to search for a putative secondary, taking possible Doppler shifts into account at both stages. We detected no companions brighter than 3% the brightness of the primary with radial velocity separations >30 km s⁻¹ from the primary. In the radial velocity separation range of 10–30 km s⁻¹ our detection limits are less robust, but we do not detect any companions brighter than 5% the brightness of RIK-210. The secondary line search is blind to companions with RV separations <10 km s⁻¹ from the primary. For reference, a velocity separation of 30 km s⁻¹ approximately corresponds to a $0.15 M_\odot$ companion at the corotation radius.

### 5.2. *Projected rotational velocity*

The projected rotational velocity was inferred from rotationally broadening the spectrum of GJ 408, an M2.5 dwarf with $v \sin i < 0.97$ km s⁻¹ (Maldonado et al. 2016), to match the spectrum of RIK-210 (Table 1). If the *K2* photometric modulation period is assumed to be the stellar rotation period and differential rotation is neglected, an expression for the stellar radius modulated by the sine of the inclination is given by:

$$R_\star \sin i = \frac{P_{\mathrm{rot}}}{2\pi} \times v \sin i. \qquad (11)$$

From this equation we find $R_\star \sin i = 1.23 \pm 0.12$ R⊙, in excellent agreement with our radius determined from the Stefan-Boltzmann law, which suggests that the stellar spin axis is nearly perpendicular to our line-of-sight.

### 5.3. *Radial velocities*

Radial velocities (RVs) were determined via cross-correlation (CCF) with RV standard stars (Nidever et al. 2002) or by using the telluric A and B absorption bands as a wavelength reference (Table 3). For RVs measured via CCF, the mean velocity was determined from cross-correlation in 6 different HIRES orders using between 3–6 (depending on the epoch) M-type RV standards, with the quoted error corresponding to the standard deviation of these 18–36 measurements. Details of telluric-based RV measurements with HIRES, and their uncertainties, are described in Chubak et al. (2012). The quoted RV errors are several times the theoretical best performance for HIRES, which is expected based on the SNR of the data and the modest $v \sin i$ of the star.

We calculated the systemic RV, $\gamma$, from a weighted mean of all measurements (Table 1). RV variations of $\gtrsim 1.5$ km s⁻¹ were observed among the HIRES spectra (Figure 14), with variations of this amplitude observed on consecutive nights in some cases. We consider the possibilities that the RV variability is due to orbital motion of a companion, rotational modulation of starspots, or some combination of the two effects.

The number of measurements is not sufficient to detect a period via Lomb-Scargle periodogram analysis, but the data are used to constrain the masses of putative companions at a range of semi-major axes (Fig. 16). From a Fisher matrix analysis of the RVs, we determined a $2\sigma$ upper limit to the mass of a putative companion on a circular orbit at the corotation radius of $m \sin i \lesssim 7.6$ M$_{\mathrm{Jup}}$.

The procedure for determining the upper limit to the mass of a putative companion is described in Boyajian et al. (2016) but we briefly summarize it here. Assuming a circular orbit, the RV measurements were folded on $4 \times 10^5$ trial orbital periods between 0.5 d and 3000 d and then fit with a sine and cosine term to represent the reflex motion and systemic RV. The Doppler semi-amplitude, $K$, and its 2-$\sigma$ uncertainty was determined for each trial period from the fit, then converted to a limit on $m \sin i$ assuming the stellar mass of RIK-210. The constraints are weakest at aliases of the observing cadence, which gives rise to the peaks in the mass limits presented in Fig. 16. More stringent mass constraints can be determined by imposing the time of mid-transit be a zero-crossing in the RVs, but given the ambiguous nature of the flux dips and the lack of photometric data concurrent with the spectroscopic observations we elected not to impose such a restriction.

In a separate analysis, we placed constraints on putative wide companions based on the maximum slope implied by the RVs. The RVs cover a period of ∼100 days, and consequently allow for the detection of a long-term trend due to the gravitational influence of a distant companion. From $10^5$ bootstrapping simulations, we determined a $3\sigma$ upper limit to the acceleration of RIK-210 of $\dot{\gamma} < 2.2$ km s$^{-1}$ yr$^{-1}$. This limit rules out the presence of additional companions more massive than 0.43 M$_\odot$ interior to 5 AU, or more massive than 0.15 M$_\odot$ interior to 3 AU. Of course, these limits do not apply interior to ∼1 AU, where the orbital period of a putative compan-ion is more than 25% the baseline of our measurements. These limits are somewhat more stringent than those implied by the Fisher matrix analysis in the relevant orbital period range.

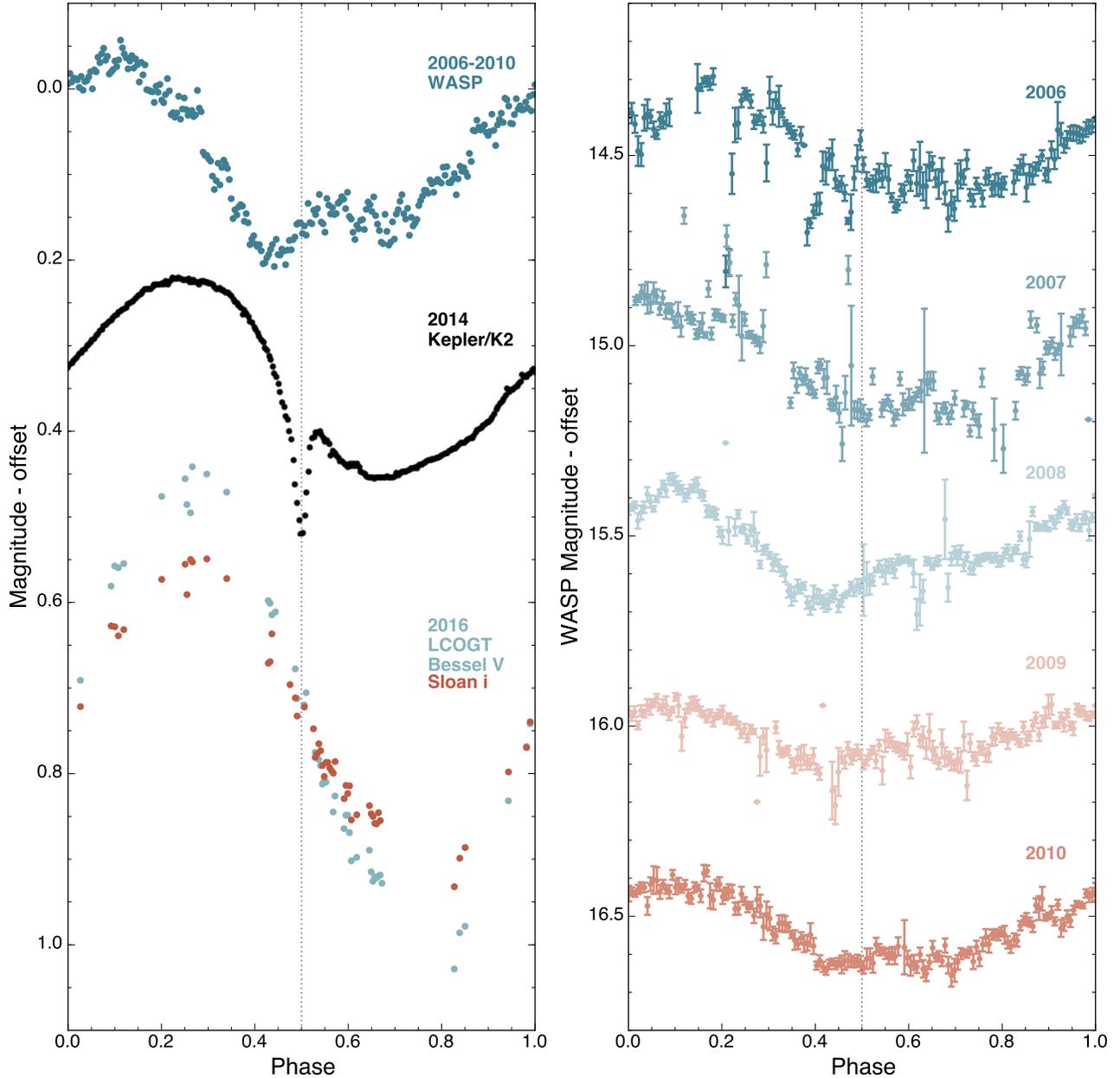

**Figure 13.** Left: Phase-binned and averaged light curves of RIK-210 at three epochs from WASP, *Kepler/K2*, and LCOGT (in both *V* and *i'* filters). Each light curve is phase-folded on the *K2* dip ephemeris. A broad depression in the WASP data between phases 0.3 and 0.5 is suggestive that the dip may have been present in previous years, but drifting in phase, becoming narrower, or both. However, inspection of the WASP data by year suggests this feature in the combined light curve may be due primarily to a combination of a questionable dip in 2006 and evolution of the spot pattern, most notable in the 2008 data. Right: WASP data ordered by year, phase-binned and averaged by weighting data points inversely to their photometric uncertainties. Data are offset vertically from the 2006 median value. Errorbars correspond to the standard deviation in a given phase bin. Only WASP data with photometric errors <0.4 mag were used. There are typically 2500–8000 measurements per year. Although there appear to be dimming events in prior years (e.g. at phases of 0.2 and 0.4 in 2006, phase 0.65 in 2008, or phase 0.45 in 2009), further inspection of the data indicate that many of the narrow structures above result from only one or two nights of observations.

Spot-induced RV modulations of $\gtrsim 1$ km s$^{-1}$ have previously been observed in weak-lined T Tauri stars, or WTTS (Prato et al. 2008; Huerta et al. 2008; Mahmud et al. 2011). The effect of rotation and spots on RVs is somewhat analogous to the Rossiter-McLaughlin effect, with the important caveats that spots are luminous and can cover a significant fraction of the star. As the most heavily spotted hemisphere rotates into view we receive more flux from the receding hemisphere and we should observe a net redshift. Conversely, as the most spotted hemisphere rotates out of view, we should observe a net blueshift. Given the highly spotted nature of RIK-210[12], it is highly likely that the dominant component of the observed RV variability is spot-induced. When phased to the rotation period, however, the RVs appear to deviate slightly from the smooth, semi-sinusoidal variation expected from spots (see e.g. Huerta et al. 2008; Mahmud et al. 2011).

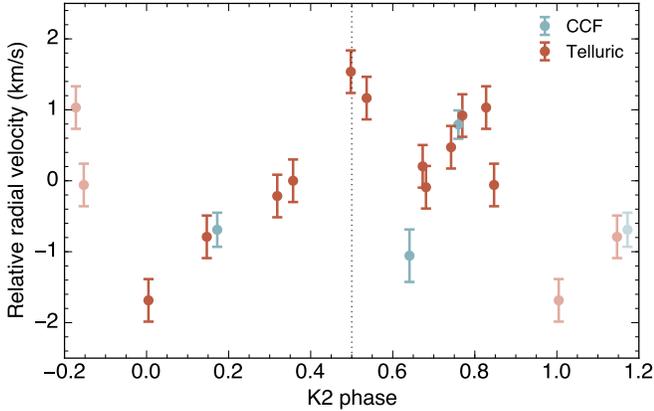

**Figure 14.** Median-subtracted radial velocities for RIK-210 phased on the period of the *K2* light curve. The dotted line indicates the approximate location of the deep dimming events present in *K2* photometry, but absent in ground-based follow-up acquired closer to the time of spectroscopic observations. For illustrative purposes we show redundant phases with lighter shaded points.

In principle, RV variability from rotation of a spotted star can interfere either constructively or destructively with the reflex motion due to a putative companion at the corotation radius, since both signals would have the same period, a scenario mentioned by van Eyken et al. (2012).

To assess whether the observed RV variability is due to stellar activity or orbital motion, we searched for wavelength-dependent RV trends. RV variations due to surface features on a star are chromatic, inducing apparent Doppler shifts only at wavelengths where the features are optically thick. Doppler shifts due to orbital motion, however, are achromatic. By performing wavelength-restricted cross correlations over 31 orders in the HIRES spectra we find a maximum slope in the RVs as a function of wavelength of $\pm 5 \times 10^{-6}$ km s$^{-1}$ Å$^{-1}$. We note that $v \sin i$ is only moderately larger than the spectral resolution of our observations, and so detecting apparent Doppler shifts due to line profile asymmetries is difficult (e.g. Desort et al. 2007).

Aigrain et al. (2012) developed a novel technique (the $FF'$ method) for calculating the expected radial velocity variability due to stellar surface features from a broadband light curve alone. We explored modeling the expected RV variability based on the phase-averaged *K2* light curve, confirming in principle that stellar RV vari-ability can mask or enhance the Doppler signal of a body at the corotation radius. We constructed model RV phase curves based on a high-order polynomial fit to the phased *K2* photometry with the flux dips removed, and subsequently fit these models to the RVs allowing the fractional spot coverage to vary. We did not find a good fit, but the $\chi^2_{\min}$ model implies a spot covering fraction of $\approx 30\%$. We note, however, that an additional free parameter in the $FF'$ models is the baseline stellar flux level in the absence of spots. This parameter is essentially unknown, but we fixed it to be the maximum of the *K2* light curve.

---

[12] Spot filling factors in late type stars of up to 50% have been inferred from high-resolution spectroscopy of molecular bands (O'Neal et al. 1998, 2004) and in the case of the young weak-line T Tauri star LkCa 4 ~80% (Gully-Santiago et al. 2016, submitted).

To avoid overinterpreting the results of the $FF'$ RV modeling, we do not present it in any more detail here and question its utility in this specific instance given (1) the uncertainty in the spot-covering fraction, (2) the lack of simultaneous photometry during the RV acquisition period, (3) the apparent change in spot modulation amplitude from follow-up photometry, and (4) the underlying requirement in the $FF'$ method that the spot-to-photosphere contrast ratio is not close to one, which may not be true for RIK-210.

Ultimately, we conclude that spots induce most but possibly not all of the observed RV variability. While the RVs do exhibit very little scatter when phased to the rotation/orbital period, they are not well fit by either a semi-sinusoidal model (i.e. a sine curve plus harmonics) as one might expect from spots or a Keplerian model. More complete phase coverage is needed to adequately model the RVs, and simultaneous photometric monitoring will greatly help in determining the contribution of spots to the RV variability.

### 5.4. Line profile variations

The chromospheric emission spectrum of RIK-210 is manifest at H$\alpha$ (exhibiting a classic double-horn profile), H$\beta$, He I 5876Å, Na I D, the Ca II H&K doublet and the triplet, and Fe I 5169 Å which are seen at all spectral epochs. All lines are variable in strength over the time series, with the exception of the Ca II triplet which remains constant. At the epoch of strongest emission, Mg Ib triplet and Fe I 5018 Å are also observed.

For the H$\alpha$ line, the measured line strengths range from $W_\alpha = -3.5$ to $-6.5$ Å. A Gaussian fit results in a 1-$\sigma$ line width of 50 km s$^{-1}$, while the full width at 10% of peak flux is 120 km s$^{-1}$. There is a clear trend between the H$\alpha$ equivalent width and rotational phase, such that stronger H$\alpha$ emission is measured when the fractional spot coverage is highest which occurs at the rotational phases where the sinusoidal component of the *K2* lightcurve is faintest (Fig. 15). Despite the significant change in line strength, there is no change over phase in the violet-to-red symmetry.

The level of activity in RIK-210 is stronger than the typical M2.5 field dwarf (e.g. Gizis et al. 2002). It is also higher than generally observed for the well-studied active star AD Leo ($W_{H\alpha} = -2.7$ to $-4.0$ Å), which was characterized as a high pressure chromosphere by Short & Doyle (1998) based on detailed radiative transfer modelling of various line species. These activity comparisons are as expected given the youth of the star.

### 6. HIGH-RESOLUTION IMAGING

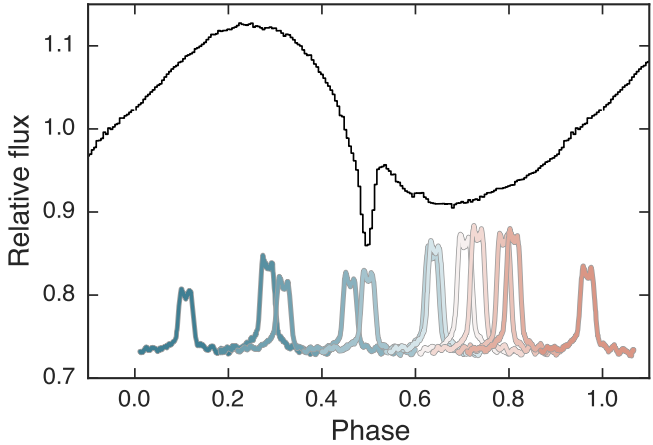

**Figure 15.** Variations of the Hα profile (colored curves) as a function of rotational phase, from the Keck-I/HIRES spectra. The black line represents the phase-averaged *K2* light curve. We emphasize that follow-up photometry, acquired around the same time as the spectra, did not show the ~20% dimming at phase = 0.5 observed by *K2*.

**Table 3**
Keck-I/HIRES radial velocities of RIK-210

| UT Date | JD | RV km s⁻¹ | σ_RV km s⁻¹ | Method | Phase |
|---|---|---|---|---|---|
| 2016-05-12 | 2457520.930 | −4.31 | 0.30 | telluric | 0.7357 |
| 2016-05-12 | 2457521.086 | −3.86 | 0.30 | CCF | 0.7632 |
| 2016-05-17 | 2457526.023 | −5.84 | 0.37 | CCF | 0.6343 |
| 2016-05-20 | 2457529.035 | −5.47 | 0.24 | CCF | 0.1656 |
| 2016-06-15 | 2457555.037 | −3.99 | 0.20 | CCF | 0.7531 |
| 2016-07-07 | 2457579.899 | −5.57 | 0.30 | telluric | 0.1395 |
| 2016-07-13 | 2457582.927 | −4.87 | 0.30 | telluric | 0.6738 |
| 2016-07-28 | 2457597.875 | −5.00 | 0.30 | telluric | 0.3110 |
| 2016-07-29 | 2457598.891 | −3.24 | 0.30 | telluric | 0.4903 |
| 2016-07-30 | 2457599.884 | −4.58 | 0.30 | telluric | 0.6655 |
| 2016-07-31 | 2457600.870 | −4.84 | 0.30 | telluric | 0.8394 |
| 2016-08-17 | 2457617.763 | −3.75 | 0.30 | telluric | 0.8198 |
| 2016-08-18 | 2457618.766 | −6.47 | 0.30 | telluric | 0.9968 |
| 2016-08-20 | 2457620.764 | −4.78 | 0.30 | telluric | 0.3493 |
| 2016-08-21 | 2457621.780 | −3.62 | 0.30 | telluric | 0.5286 |

Speckle imaging observations of RIK-210 were acquired at Gemini South Observatory with the DSSI instrument (P.I. Steve Howell). The speckle observations revealed no companions with $\Delta m \lesssim 4$ mag at separations between 0.1–1.37″. Beyond 0.5″, we obtained more stringent contrast limits of $\Delta m \lesssim 4.8$ mag at 692 nm, or $\Delta m \lesssim 5.4$ mag at 880 nm.

We also obtained near-IR adaptive optics (AO) imaging for RIK-210 on UT 2016 Jul 17 using NIRC2 (P.I. Keith Matthews) with the Keck II Natural Guide Star (NGS) AO system (Wizinowich et al. 2000). We used the narrow camera setting with a plate scale of 10 mas pixel⁻¹. This setting provides a fine spatial sampling of the instrument point spread function. The observing conditions were good with seeing of 0″.5. RIK-210 was observed at an airmass of 1.29. We used the $K_S$ filter to acquire images with a 3-point dither method. At each dither position, we took an exposure of 0.25 second per coadd and 10 coadds. The total on-source integration time was 7.5 sec.

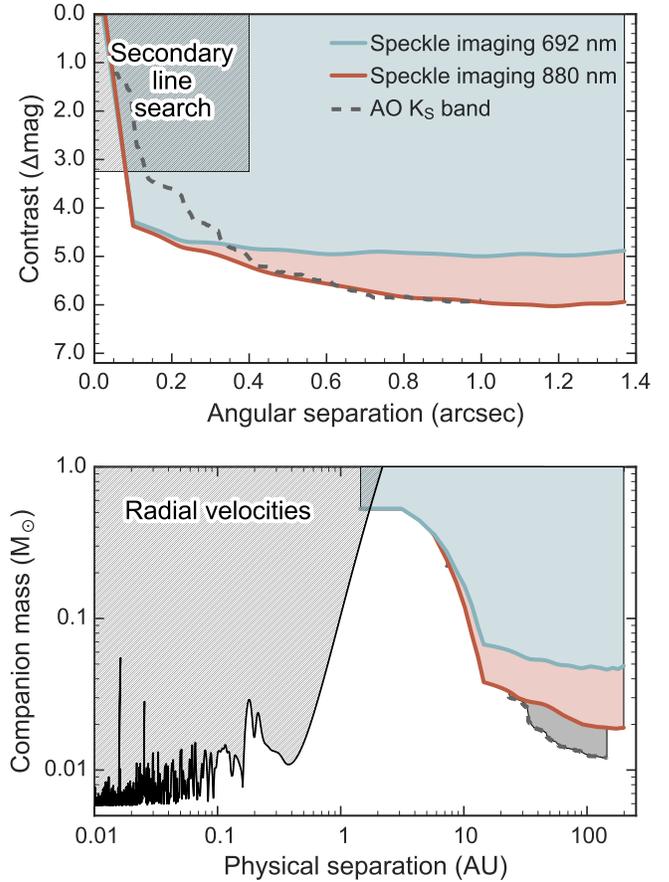

**Figure 16.** Constraints on the brightness (top panel) or mass (lower panel) of putative companions to RIK-210 from optical speckle imaging (blue and red regions), NIRC2 adaptive optics imaging (grey dashed line in top panel, grey region in bottom panel), a HIRES secondary spectral line search, and the RV time series.

The raw NIRC2 data were processed using standard techniques to replace bad pixels, flat-field, subtract thermal background, align and co-add frames. We did not find any nearby companions or background sources at the 5-σ level. We calculated the 5-σ detection limit following Wang et al. (2014). We defined a series of concentric annuli centered on the star. For the concentric annuli, we calculated the median and the standard deviation of flux for pixels within these annuli. We used the value of five times the standard deviation above the median as the 5-σ detection limit. The 5-σ detection limits are 2.1 mag, 3.6 mag, 5.4 mag, and 5.9 mag for 0″.1, 0″.2, 0″.5, and 1″.0, respectively.

We combined our high-resolution imaging constraints with the constraints from the secondary spectral line search and lack of a significant RV trend to place limits on putative companions to RIK-210. The combined constraints are depicted in Figure 16. To convert imaging constraints from Δ mag to companion mass we used the pre-main sequence evolutionary models of Baraffe et al. (2015), using the tabulated *R* and *I* magnitudes as proxies for the 692 nm and 880 nm speckle imaging constraints, respectively. In particular, the *I* band constraints rule out companions down to 20–25 $M_{\rm Jup}$ between 60–200 AU, and 25–40 $M_{\rm Jup}$ between 15–60 AU.

## 7. DISCUSSION

Here we discuss possible explanations for the K2 light curve of RIK-210, in light of its previous and subsequent evolution as observed with WASP and LCOGT, and the spectroscopic monitoring undertaken with Keck-I/HIRES. We consider three general classes of explanations: rotational modulation of stellar surface features, co-rotating circumstellar gas and dust, and phenomena associated with a young, planetary system. In Table 4 we present a broad overview of the most developed theories.

### 7.1. Rotational modulation of high-latitude spots?

Given the youth of RIK-210 and the large amplitude semi-sinusoidal variability, it is evident that the star is heavily spotted. As the dimming events are essentially in phase with the rotation period of the star, we consider the possibility that the events are due to a spot pattern on the stellar surface. The main difficulties facing a spot-based explanation are that the dimming events are both deep and brief relative to the rotation period. Variability due to any feature on the stellar photosphere typically endures for half of the rotation period, while the dimming events in RIK-210 are always <15% of the rotational phase.

In principle, a surface feature can rotate into view for less than half of the rotation period if the feature is confined to high latitudes and the stellar rotation axis is tilted modestly away from an equator-on orientation. Thus, a cool spot near the poles, if the rotation axis is tilted slightly away from the observer, could produce a dimming event that endures for less than half the rotation period. Indeed, the existence of polar spots on young, rapidly rotating stars has both theoretical (e.g. Schuessler & Solanki 1992; Granzer et al. 2000; Yadav et al. 2015) and observational support (e.g. Hatzes 1995). However, in this scenario it is still difficult to produce sudden or "angular" flux dips because spots are subjected to limb darkening and foreshortening.

The K2 observations seem to be in tension with a starspot explanation, given how rapidly the depth, duration, and general morphology changes from epoch to epoch. Meanwhile, the globally averaged spot distribution appears to remain stable throughout the K2 campaign and in fact the follow-up LCOGT photometry confirms the ephemeris and general morphology of the disk-averaged spot pattern more than two years later. Moreover, spots or spot groups on T Tauri stars have measured lifetimes of years or even decades (e.g. Herbst et al. 1994; Bradshaw & Hartigan 2014).

The maximum depth of the dimming events is ≈18%. The depth of a dimming event due to starspots is determined by both the opacity and size of the spot(s) at the observed wavelength. The minimum spot size required to produce a given flux decrement is set by the limiting case of a completely opaque spot. In this scenario, the spot size is approximated in the same manner as the size of a transiting planet. Namely, $R_{\rm spot}/R_* = \sqrt{\delta}$, where $\delta$ is the depth of the dimming event. The deepest dimming events observed in RIK-210 would require a *minimum* spot or spot group size of $\sim 0.4 R_*$ or $\sim 0.5 R_\odot$. This minimum size is so large that it likely could not be confined to the restricted range of latitudes required to get dimming events so brief relative to the rotation period.

Furthermore, spots are not opaque and the observed contrasts around M-type stars correspond to temperature differences of only a few hundred Kelvin (Andersen & Korhonen 2015).

To have a spot or spot group of sufficient size, contrast, and latitude to produce dimming events of $5-15\%$ for only $\sim 15\%$ of the star's rotational phase presents a fine-tuning problem, and is perhaps physically impossible. We conclude that, despite being synchronized with the stellar rotation, the dimming events observed by K2 are unlikely to be related to spot behavior on the star.

### 7.2. Eclipses of an accretion hotspot?

Although there is no evidence for ongoing accretion analogous to that of classical T Tauri stars, given its youth, RIK-210 could still be experiencing residual gas infall from gaseous and dusty debris in an only recently dispersed circumstellar disk. A star undergoing accretion may develop a hotspot on its surface, as infalling material is deposited onto the star. If the material falls in along magnetic field lines, it should be deposited near one of the magnetic poles. If a star has an accretion hotspot on its surface, the disk averaged brightness will decrease when the hotspot rotates out of view (see e.g. Romanova et al. 2004). Moreover, the magnetic poles need not be coincident with the rotational poles. As such, even a star with its rotational axis aligned perpendicular to the observer's line-of-sight could host an extended hotspot that is in view for more than half of the rotational phase. This scenario is analogous to the cool starspot scenario, and was considered as a possible explanation for the dimming events seen in PTFO 8-8695 (Yu et al. 2015).

Following the methodology of those authors, one can estimate the mass accretion rate from the depth of flux decrements:

$$L_{\rm acc} = \frac{GM_* \dot{M}}{R_*} \sim \delta L_{\rm bol}, \qquad (12)$$

where $\dot{M}$ is the mass accretion rate and $\delta$ is the depth of dimming events. For dimming events of $\sim 9$–18% and the derived stellar parameters in Table 1, the scenario above would require mass accretion rates of $\dot{M} \sim 1$–$3 \times 10^{-9} M_\odot$ yr$^{-1}$. Despite the modest burst (with flare-like decays) observed in the K2 light curve, we consider this scenario unlikely as the star lacks both spectroscopic accretion indicators (the enhanced hydrogen emission observed from the spectra is consistent with chromospheric emission) and the infrared excess that would be associated with an inner disk.

### 7.3. Eclipses of prominences?

Some rapidly rotating stars may be surrounded by dense clouds of partially ionized gas (so-called slingshot prominences) at a distance of several stellar radii and co-rotating with the stellar magnetosphere. These clouds may have projected areas as large as 20% of the stellar disk (Collier Cameron & Robinson 1989a,b; Collier Cameron et al. 1990). While such clouds have been observed from absorption in the Balmer and Ca II H & K lines, there is no appreciable continuum absorption. Thus, transits of such clouds are not thought capable of producing the deep dips observed in RIK-210. However, if a large prominence contributes significant optical flux relative to the star via Paschen-continuum bound-free emission, it may be possible to produce broadband optical dimmings of a few percent when such a prominence passes *behind* the star. Nevertheless, the dip depths observed in RIK-210 are significantly deeper than a few percent. We conclude that eclipses of slingshot prominences can not produce the 20% dimmings observed in the *Kepler* bandpass.

**Table 4**
Proposed explanations for the transient transits of RIK-210

| Scenario | Supporting evidence | Conflicting evidence | Plausibility |
|---|---|---|---|
| eclipsing binary or transiting brown dwarf | deep, V-shaped dimmings | inconsistent with RVs, archival/follow-up photometry | ruled out |
| high-latitude star spot | synchronicity between rotation and dip periods, spotted star | inter-rotation variations and combination of deep depths with short durations | highly unlikely |
| high-latitude accretion hot spot | synchronicity between rotation and dip periods, modest photometric burst | lack of spectroscopic accretion indicators or IR excess associated with inner disk | highly unlikely |
| eclipses of prominences | synchronicity between rotation and dip periods, magnetically active star | depths are too deep | highly unlikely |
| transits of magnetospheric clouds | synchronicity between rotation and dip periods, magnetically active star | does not readily explain the shallow, short-duration flux dips | most likely |
| dipper | synchronicity between rotation and dip periods, depths and variable morphologies of dimming events | lack of spectroscopic accretion indicators or IR excess associated with inner disk | highly unlikely |
| transits of an enshrouded protoplanet | RV variability difficult to explain with spots alone | requires extended dusty tail or swarm of satellites to explain long durations, in tension with archival/follow-up photometry, synchronicity between rotation and dip periods not a requirement | somewhat plausible |
| tidal disruption of planetary or cometary material | variable depths, durations, and morphologies of dimming events, may explain shallow, short-duration dips that appear to drift in phase | does not readily explain synchronicity between rotation and dip periods, requires high eccentricities and/or very low densities of orbiting bodies | somewhat plausible |

### 7.4. *Transits of magnetospheric clouds?*

Transits by circumstellar, magnetospheric clouds have been suggested as a source of photometric variability for high-mass stars (Groote & Hunger 1982). In particular, the magnetic Be star $\sigma$ Ori E shows clear, $\sim$5% eclipses twice per rotation period; these eclipses have been attributed to clouds of plasma which are trapped in the magnetosphere and most dense at the corotation radius (Townsend et al. 2013). The phenomenon of a narrow flux dip in phase with, and superposed on, semi-sinusoidal modulation has also been observed in the X-ray light curve of the accreting white dwarf PQ Gem (Mason 1997; Evans et al. 2006). In that case, the authors attribute the dip to matter from an accretion disk being lifted up out of the plane and into the line-of-sight by the leading edge of the magnetic field lines.

Fully-convective, low-mass pre-main sequence stars have magnetic field strengths typically in the 0.1–1 kG range (e.g. Johns-Krull 2007; Donati & Landstreet 2009). Charged material in orbit about the star may become trapped in the magnetosphere at the corotation radius, where closed field lines thread the orbital plane. Magnetized stars may also accrete matter from a circumstellar disk via so-called funnel flows (e.g. Romanova et al.

2002). Such accretion columns, if they contain dust and transit the disk of the star, could lead to optical extinction. Indeed, this idea underpins some explanations for the dipper phenomenon discussed below.

Whether the magnetically entrained material is a cloud of plasma analogous to those observed in high-mass stars, or a dusty accretion column, this model could naturally explain the synchronicity between the rotation period and the dimming events, as the field lines, and thus flows, thread the equatorial plane at corotation. Since the accretion timescale is expected to be of order the free-fall timescale, i.e. much shorter than the orbital period, this model might explain the variable depths and morphologies of dimming events.

The most challenging observation to the accretion column scenario is that RIK-210 lacks spectroscopic accretion indicators. Perhaps we are witnessing the low-level, end state of accretion for a post T Tauri star. Calculations below show that only a modest amount of dust is needed to produce the deep dimming events observed with *K2*. The dust would have to be charged, or coupled to plasma, in order to be dragged by the field lines. The source of such dust close to the star remains a mystery. Moreover, the transient nature of the dimming events would seem to suggest that the dust must be episodically replenished. The modest amount of dust remaining in the system at AU scales may spiral in via P-R drag, but if it is charged one might expect it to be captured by the magnetosphere and subjected to an outward magnetocentrifugal force prior to reaching the inner magnetospheric regions. Regardless of the actual source of the dust, if this model is correct, it reveals a previously unappreciated aspect of the early environments of close-in exoplanets.

We consider this theory quite plausible for the variability observed in RIK-210, however several unanswered questions remain: in the absence of a tilt between the rotation and magnetic axes, how does the cloud remain so compact as to produce dips that are so narrow in rotational phase? How does one explain the inter-rotation variability in morphology, depth, and duration? Perhaps most importantly, the magnetospheric model struggles to explain the short-duration, shallow dips or at least implies that accretion of the material is clumpy rather than smooth.

### 7.5. *Obscuration by a circumstellar disk?*

Variable obscuration due to a circumstellar disk has long been an explanation for the photometric variability of some T Tauri stars. For example, UX Ori is the prototype of what Herbst et al. (1994) described as Class III variability within early type T Tauri stars. At lower stellar masses, AA Tau is the archetype for similar variability, which has been attributed to a warped inner disk edge interacting with the stellar magnetosphere (Bouvier et al. 2003, and references therein).

Multi-wavelength observations of this phenomenon in Orion support this hypothesis, showing that the dimming events are deeper in the optical and shallower in the infrared, as one would expect from extinction by dust (Morales-Calderón et al. 2011). The advent of high cadence, precision photometry using space telescopes has revealed an unprecedented level of detail to this general class of variable young stars now called "dippers" (Cody et al. 2014; Stauffer et al. 2015; Ansdell et al. 2016).

In some cases where the stellar rotation is evident, such as the case of Mon 21, the dimming events have been observed to be synchronized with the stellar rotation period (Stauffer et al. 2015). The coincidence between stellar rotation and the orbital period of transiting debris in this model is a natural consequence of disk-locking, where the inner disk edge is truncated near the corotation radius due to interactions with the stellar magnetosphere.

Dippers also exhibit variable depths, but often so variable that the dimming events disappear entirely (EPIC 204137184 is one such quasi-periodic example in Upper Sco). Ansdell et al. (2016) classified the main morphological types of dimming events seen in Upper Sco dippers as: symmetric, trailing tail, and complex. While some of the dimming events in RIK-210 might be classified as symmetric or complex, noticeably absent in that list is the leading tail morphological class, which seems most appropriate for RIK-210.

While some dippers show strong spectroscopic accretion signatures from shocked gas falling onto the star, many exhibit only weak accretion indicators such as those observed in RIK-210. However, within Upper Sco, dipper stars typically have infrared excesses consistent with a protoplanetary disk extending in to ~10 $R_*$ (Ansdell et al. 2016). By contrast, RIK-210 has no evidence of an inner disk, yet the obscuring material, if in orbit, rests at a distance of ~9 $R_*$. The essential differences between RIK-210 and young dipper stars are that the dimming events in RIK-210 are more constant in depth and shorter in duration, while RIK-210 also lacks evidence for an inner disk.

For the reasons discussed above, it is extremely unlikely that RIK-210 belongs to the class of young dipper stars. However, as we can not completely exclude this scenario, we note that if RIK-210 is indeed a dipper then it would be the first such object with (1) no inner disk, (2) a light curve dominated by rotational modulation of starspots rather than by dimming events, and (3) a relatively high degree of both periodicity and morphological consistency between dimming events.

### 7.6. *A co-rotating dust component?*

In addition to gas phenomena, dust in the vicinity of the stellar co-rotation radius could be responsible for the narrow dips in RIK-210. Dimming events over ~15% of the rotational phase could be explained by an extended structure of dust obscuring the star and scattering starlight away from the line-of-sight. If this is the case, the existence of such a significant amount of dust close to the star must be explained. For example, the dust could be primordial in origin (i.e. remnants of the late stages of planet formation), the product of a recent major collision, or from sublimation of volatile-rich bodies.

By equating the stellar luminosity with the luminosity of a dust disk extending out to the corotation radius[13], the temperature of dust at the corotation radius is estimated by,

$$T_{\rm corot} = 0.34\, T_{\rm eff} \left( \frac{P}{1\,\rm day} \right)^{-1/3} \left( \frac{R_*}{R_\odot} \right)^{1/2} \left( \frac{M_*}{M_\odot} \right)^{-1/6}. \tag{13}$$

Using the Kobayashi et al. (2011) sublimation temperatures for various grain compositions, we inverted the above equation to find the minimum orbital period at which dust of a given composition could exist in the solid phase (Fig. 17). The transiting debris around RIK-210 is safely outside the dust sublimation radius for olivine, pyroxene, obsidian, and pure carbon, though it is close to the sublimation radius for pure iron. Note that this analysis assumes a circular orbit.

Dust in orbit about the star will be subjected to radiation pressure and Poynting-Robertson (P-R) drag. The minimum size of dust grains that can survive on circumplanetary orbits is set by the balance of these forces (Burns et al. 1979; Kennedy & Wyatt 2011).

From Lecavelier Des Etangs et al. (1999), the ratio of radiation pressure to the gravitational force is given by:

$$\beta = 0.2 \left( \frac{L_*/L_\odot}{M_*/M_\odot} \right) \left( \frac{s}{1\,\mu\rm m} \right)^{-1}, \tag{14}$$

where $s$ is the grain size. For RIK-210, equilibrium between gravity and radiation pressure is established for grain sizes of 0.075 $\mu$m. For smaller grains, radiation pressure dominates, and for larger grains, gravity dominates.

From Burns et al. (1979), the timescale for a dust grain to spiral in to the star via P-R drag is,

$$t_{\rm PR}(r) = \frac{cr^2}{4GM_*\beta}. \tag{15}$$

For the parameters of RIK-210, the P-R drag timescale is approximately 2.5 years for 0.1 $\mu$m grains or 25 years for 1.0 $\mu$m grains. Thus, if the *K2* extinction events were due to dust grains of 0.1 $\mu$m in size or smaller, it is plausible that the majority of this dust has since spiraled into the star and thus explain why no such dimming events were detected with follow-up photometry.

---

[13] We note this equation neglects scattering, self-shielding, or heating by accretion. See Bodman et al. (2016) for details.

One can calculate to first order the mass of dust, $M_{dust}$, from the maximum depth of a dimming event. From Lecavelier Des Etangs et al. (1999), the maximum extinction due to a cloud of grains is given by,

$$\left(\frac{F_{ext}}{F_*}\right)_{max} = \exp\left(-\frac{6 M_{dust} \int s^2 dn(s)}{4\pi\rho R_*^2 \int s^3 dn(s)}\right), \quad (16)$$

where $s$ is the grain size, $dn(s)$ is an analytic grain size distribution, and $\rho$ is the dust density. For the grain size distribution adopted by Lecavelier Des Etangs et al. (1999) and a grain density of 3 g cc$^{-1}$, we find that $\sim 10^{-9} M_\oplus$ could extinct RIK-210 by 20%.

Optical attenuation by a transiting dust cloud is primarily the result of scattering of light away from the line-of-sight, rather than absorption. Consequently, at particular orbital phases, a dust cloud may forward scatter starlight into the line-of-sight and contribute observable excess flux to the light curve (Rappaport et al. 2012). For RIK-210, we do not see obvious evidence for forward scattering. However, it is difficult to draw conclusions because such evidence critically depends on how well the stellar continuum level is fit. For example, transits 8 and 10 appear to show modest post-egress flux excesses evocative of forward scattering (see Fig. 4), though these features may also be a result of improper continuum fitting. For disintegrating *Kepler* planets on ultra-short-period orbits, the effect of forward scattering is miniscule but statistically significant because of the sheer number of transits over a much longer time baseline.

If there is a significant amount of dust close to the star, the dust will be heated by the star and re-radiate in the near-IR. In § 2.2 we showed that RIK-210 lacks warm dust, but there is marginal evidence for cool dust at AU scales. Thus, if corotating dust is responsible for the dimming events, it must be scarce enough to avoid producing a detectable near-IR excess. We estimated the dust-to-star flux ratio at 24 $\mu$m, where a putative excess would be most easily be detected, using the following equation:

$$\frac{F_{24,dust}}{F_{24,*}} = \frac{A_{dust}}{A_*}\frac{T_{dust}}{T_*}, \quad (17)$$

where $A_{dust}$ is the total cross section of dust grains in the obscuring region and $T_{dust}$ is the dust equilibrium temperature at corotation, here taken to be $\approx$830 K. The implied excess depends on the number of dust grains and their typical geometric cross section. The number of grains, in turn, is highly sensitive to the assumed geometrical distribution. The total cross section of dust is estimated by $A_{dust} = \langle\tau\rangle A_{geom} = N(\pi s^2)$, where $\langle\tau\rangle \approx 0.15$ is the average optical depth, $A_{geom}$ is the geometrical area of the dusty region, $N$ is the number of grains, and $s$ is the typical grain size. For 0.1 mm-sized grains, the implied 24 $\mu$m dust-to-star flux ratio is $\sim$0.3 assuming the dust is distributed over a thin cylinder centered around the star with radius equal to the corotation radius and height twice the radius of the star. However, assuming the same dust properties, the flux ratio could be doubled if the dust were spread out into an annulus at the corotation radius, or an order of magnitude lower if the dust were arranged in a thin strip as one might expect from an accretion column. Thus, given the uncertainties in dust properties and the geometric distribution of dust in the system, it is difficult to know whether the proposed amount of dust would have been detected in the near-IR. However, we argue there are many plausible geometries and dust properties for which such an excess would have evaded detection.

Dust provides a convenient explanation for the dimming events in that (1) only a modest amount is required to produce deep dimming events, and (2) it can be evacuated somewhat rapidly depending on its size and composition. However, as mentioned at the beginning of the section, the source of such dust must still be explained, given that the star apparently lacks an inner disk. One possibility is a giant impact-type collision between oligarchs.

Some theories predict that collisions between oligarchs on closely packed orbits are a common occurrence in the late stages of planet formation. For example, Chiang & Laughlin (2013) showed that newly-formed, close-in super-Earths lack the energy to scatter each other out of the parent star's gravity well, so planet-planet scattering events instead lead to mergers. Perhaps the dimming events observed in *K2* are the result of a recent collision between two or more protoplanets, which could release a profuse amount of planetary debris. Depending on the energy of the collision, some fraction of material might escape the gravity of the protoplanets, while the remaining material is retained within the combined Hill spheres of the protoplanets. We note that scenarios invoking a collision may be inherently unlikely based on timescale considerations.

Another possible source for dust is a comet or family of comets. The idea of "falling evaporating bodies" has been studied extensively, particularly in the context of the young star $\beta$ Pic, which exhibits spectroscopic peculiarities ascribed to evaporating exocomets (e.g. Kiefer et al. 2014, , and references therein). As we established, the required mass to create such deep dimming events is quite modest, comparable to the mass of Hale-Bopp. What could bring cometary material onto such a short period orbit around the star? One prediction of disk-driven migration is that a migrating planet will trap planetesimals (or comets) in mean motion resonance, allowing them to reach stargrazing orbits (Quillen & Holman 2000; Thébault & Beust 2001).

### 7.7. Boil-off of a protoplanet atmosphere?

Owen & Wu (2016) predicted that a highly inflated protoplanet newly exposed to vacuum conditions after the confining pressure of the protoplanetary disk has gone will experience profuse atmospheric mass loss via a Parker wind. This mass loss is catalyzed by stellar continuum radiation, as opposed to the EUV/X-ray-driven photoevaporative mass loss that becomes important at later times. Those authors suggest extraordinary mass loss rates of $\sim$0.01 M$_\oplus$ yr$^{-1}$ may be possible over characteristic timescales of $\mathcal{O}(10^3$ yr). Clearly, this scenario presents a fine-tuning problem as it requires our observations to be coincident with this relatively short-lived

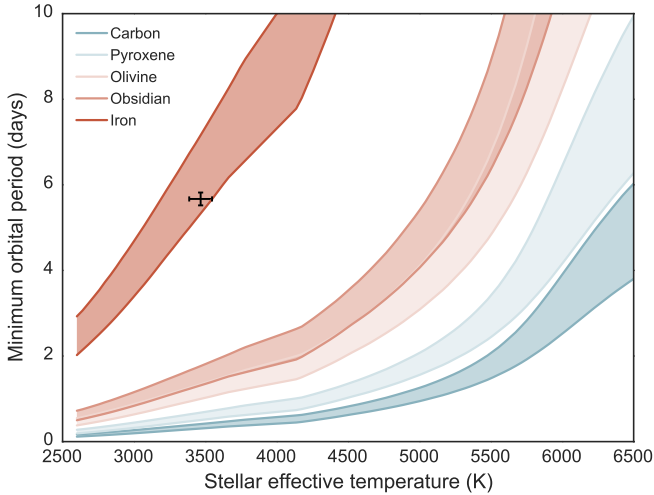

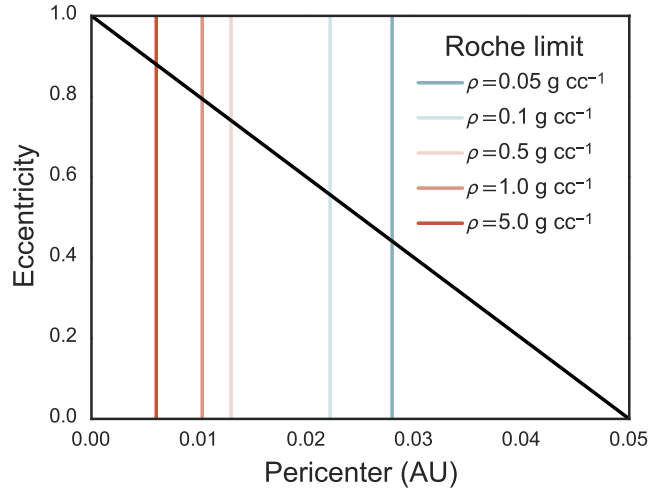

**Figure 17.** Minimum orbital period for grains of varying compositions to exist in the solid phase, as a function of the central star's effective temperature. Dust sublimation temperatures are taken from Kobayashi et al. (2011), and the stellar parameters needed to calculate the corresponding orbital periods are obtained from the PARSEC v1.2s pre-main sequence models (Bressan et al. 2012; Chen et al. 2014). For each grain composition, the upper and lower boundaries of the shaded regions are set by the 5 and 10 Myr isochrones, respectively. The position of RIK-210 is shown by the black point. Dust grains with compositions ranging from carbon to obsidian could exist in the solid phase at the corotation radius, while iron grains may sublimate.

phase of a planet's evolution.

### 7.8. *Tidal disruption of a planet?*

The leading tail morphology of the RIK-210 dimming events is somewhat suggestive of an orbiting body experiencing Roche lobe overflow. For a body in a circular orbit, the minimum period allowed before tidal disruption is given by equation 2 from Rappaport et al. (2013):

$$P_{min} \simeq 12.6 \text{ hr} \left( \frac{\rho_p}{1 \text{ g cm}^{-3}} \right)^{-1/2}. \quad (18)$$

For typical planetary densities observed in the solar system ($\sim$0.5–5 g cm$^{-1}$), this minimum period is on the order of hours, i.e. much shorter than the period of dimming events around RIK-210. In fact, for a body to undergo tidal disruption in a circular orbit of period 5.67 days would require a density of $\lesssim 0.01$ g cc$^{-1}$. This is an order of magnitude lower than the least dense bodies in our solar system, including comets. Planets are presumed to be less dense at young ages, as they are still undergoing Kelvin-Helmholtz contraction, but models of e.g. Jovian-mass planets at an age of 5–10 Myr still predict densities that are $\sim$0.5 g cc$^{-1}$ (Baraffe et al. 2003; Spiegel & Burrows 2012).

However, it is possible that a body or collection of bodies are in an eccentric orbit, with a close pericenter passage to the star. In this scenario, the body or bodies undergo periodic tidal disruption upon each crossing of the Roche radius, analogous to the disruption of comets in the solar system. The pericenter distance for an orbit of semi-major axis $a$ and eccentricity $e$ is given by,

$$a_{peri} = a(1-e). \quad (19)$$

**Figure 18.** Pericenter separation for a range of eccentricities assuming a semi-major axis of 0.05 AU (black line). The colored lines reflect the limiting Roche radius for bodies of varying densities. Eccentricities greater than 0.4 would be required to bring a body interior to its Roche limit for any of the densities plotted here.

The Roche limit for a self-gravitating, incompressible fluid satellite is given by,

$$a_{Roche} \approx 2.44 R_* \left( \frac{\rho_*}{\rho_s} \right)^{1/3}, \quad (20)$$

where $R_*$ and $\rho_*$ are the radius and density of the star, respectively, and $\rho_s$ is the density of the satellite.

By setting the Roche limit equal to pericenter, we can estimate the minimum eccentricity needed to bring a planet of a given density interior to its Roche limit (Fig. 18). For planet densities >0.1 g cc$^{-1}$, an eccentricity $\gtrsim$0.6 is required. We note that planets that are *both* massive and highly eccentric are seemingly disfavored by the RV measurements which do not exhibit a cusp or variability above the 2 km s$^{-1}$ level.

An appealing aspect of this scenario is that it might naturally explain the changing morphology between consecutive dimming events, as the orbiting body or bodies are disrupted upon each pericenter passage. It might also explain the apparent vanishing of dimming events in follow-up photometry. Complete disruption of a body or bodies could result in a ring of material around the star which would then be subject to radiation pressure and P-R drag. Nevertheless, we consider this scenario unlikely given the high eccentricities and/or low satellite densities required.

### 7.9. *Transits of an enshrouded protoplanet?*

Any planetary explanation for the dimming events faces the challenge of accounting for the depths of the events. The loss of light during a transit or eclipse depends on both the size ratio and brightness ratio of the occulting body to the central star. However, the secondary line search and radial velocity time series suggest it is quite unlikely that RIK-210 hosts a companion massive enough to contribute significantly to the total flux received from the system. Furthermore, we find no evidence of secondary eclipses within the *K2* photometry. The facts that the dimming events are deep and longer than the expected transit duration for a point mass suggest an occulting body (or collection of bodies) that is large relative to the star. On the other hand, the fact that the events are V-shaped, as opposed to flat-bottomed, suggests that if the occulting body is completely opaque, it cannot have *both* an impact parameter close to zero and a size larger than the star itself, else it would produce a total eclipse. Of course, an enshrouded protoplanet need not necessarily fill its Hill sphere (Kennedy & Wyatt 2011), nor would the material filling such a volume be completely opaque.

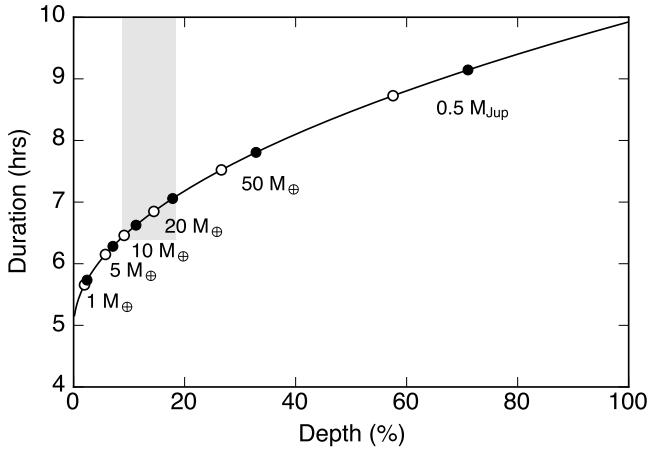

**Figure 19.** Expected depths and durations for equatorial transits of RIK-210 by an optically thick Hill sphere for planets of various masses. Fiducial masses are plotted as black points and annotated with text. However, the planet mass and fraction of the Hill sphere that is optically thick is completely degenerate in this plane. Thus, we show the effect of how these fiducial planet masses move along this curve when only 90% of the Hill sphere is optically thick (open circles). The grey shaded region indicates the range of depths and durations observed in the *K2* light curve.

The Hill radius is approximately given by:

$$r_{\rm Hill} \approx a(1-e)\left(\frac{m_p}{M_*}\right)^{1/3},\qquad(21)$$

where $a$ is the semi-major axis, $e$ is the eccentricity, and $m_p/M_*$ is the mass ratio between the planet and star. If a planet were enshrouded in a dusty envelope, the Hill radius sets the maximum extent of that envelope.

For planets in the mass range of 1 $M_\oplus$–10 $M_{\rm Jup}$ orbiting at the corotation radius of RIK-210, the Hill radius is in the range of 0.2–2.8 $R_\odot$. We calculated the transit durations of objects in this size range by inverting eqn. 8. We find transit durations of 5.8–16.6 hrs, for an inclination of $i = 90°$ and impact parameter of $b = 0$. Of course, if the enshrouded planet were at the high end of the mass range above, the Hill radius would be larger than the star itself, and the orientation considered above would result in a total eclipse of the star if the sphere were totally opaque.

A challenge presented to this scenario is accounting for both the depths and durations simultaneously. The maximum durations observed are so long as to require an object with a size larger than the star itself. For example, the Hill radius of a $10 M_{\rm Jup}$ planet is approximately $2.8 R_\odot$. If transiting the equator of the stellar disk such an object could lead to the maximum observed dimming duration of approximately 16 hours.

In Figure 19 we show the expected transit depths and durations of an optically thick Hill sphere in orbit at the corotation radius of RIK-210. Points on the curve are degenerate in the parameters of planet mass and the fraction of the Hill sphere that is optically thick. Still, the figure is useful for determining a lower limit to the mass of a hypothetical transiting protoplanet in the scenario described above. In the limit of a completely opaque Hill sphere, a planet mass above ∼10 $M_\oplus$ is required. At high planet masses, a meaningful limit can not be placed since the fraction of the Hill sphere filled with optically thick material can be arbitrarily low. We note that if material were escaping a putative protoplanet, it could of course occupy a larger volume than the Hill sphere and thus lower planet masses may be plausible. In any event, this model can not explain the long durations of the dimming events. An extended tail of optically thick material could be invoked to reconcile this matter.

In this scenario, one must still explain the existence of such a dusty envelope, and also the changing morphologies between transits. If an enshrouded planet is surrounded by a highly non-uniform distribution of dusty material or satellites, perhaps gravitational interactions between the planet and clumps or satellites are reconfiguring the spatial distribution of the bodies on timescales shorter than its orbit.

There is now a wealth of literature on circumplanetary swarms of dust and debris. Current research on this topic is mainly theoretical or the result of numerical simulations and much of it is motivated by the directly imaged planet Fomalhaut b, which is apparently visible in the optical but not the infrared due to scattering by a proposed cloud of dust surrounding the planet (Kalas et al. 2008). Kennedy & Wyatt (2011) predicted that circumplanetary swarms of irregular satellites may be common around young exoplanets, and proposed this as an explanation for Fomalhaut b. Nesvorný et al. (2007) studied the capture of irregular satellites by Jovian planets migrating within a planetesimal disk, finding many hundreds of such satellites could be captured early in the planet formation process. Additionally, the directly imaged protoplanet LkCa15 b (Kraus & Ireland 2012; Thalmann et al. 2016) appears to exhibit a tail-like structure, though we note that star is much younger and the protoplanet resides on a much wider orbit.

Finally, we note that the Darwin instability (Darwin 1879) is a concern for corotational orbits. This occurs when the rotational angular momentum of the primary exceeds one third the orbital angular momentum of the secondary, ultimately leading to spiral-in over the tidal decay timescale. We find for RIK-210 that bodies above a few Earth masses are Darwin stable in a corotational orbit.

### 7.10. *Significance of the corotation radius*

Conspicuously, the dimming events observed in RIK-210 are essentially synchronous with the stellar rotation. Above, we discussed why it is extremely unlikely that these dimming events could be due to anything on the stellar surface from empirical considerations. Here we provide context, in the form of recent observational and theoretical work, for why planets or planetary material might be *expected* to be found near the corotation radius around stars with ages approximately equal to the disk dispersal time ($\lesssim$10 Myr). At these ages, the stellar rotation period reasonably tracks the expected location of the inner disk edge. At older ages, the star first spins up as it contracts on to the main sequence (bringing the corotation radius inwards), then spins down as it evolves on and away from the main sequence (causing the corotation radius to be moved outwards).

In a protoplanetary disk, a low-density magnetospheric gap extends from the stellar surface to the inner disk edge. Theoretical investigations and numerical simulations of accretion disks around magnetized stars find the inner disk edge is truncated at the Alfvén radius, where the magnetic energy density in the disk is equivalent to the kinetic energy density. In practice, for T Tauri stars, the Alfvén radius is approximately 20-30% interior to the corotation radius (Long et al. 2005).

Romanova & Lovelace (2006) found the effect of the magnetospheric gap is to greatly reduce the rate of planetary migration through disk interactions. Papaloizou (2007) also studied the orbital evolution of a planet that has migrated through the disk into the magnetospheric cavity, finding that protoplanet interactions with the stellar magnetosphere should not result in significant orbital evolution after entering the gap. There is also observational support for truncation of the disk at corotation. Meng et al. (2016) recently reported the first measurement of the inner edge of a protoplanetary disk using photo-reverberation mapping. Those authors found a value for the disk edge that is consistent with expectations for the corotation radius.

Of the confirmed or candidate exoplanets around T Tauri stars discovered to date, all are on orbits near the presumed corotation radius. David et al. (2016b) pointed out that K2-33 b orbiting slightly interior to the corotation radius, with $P_{orb}/P_{rot} \approx 0.86$. The period of the candidate transiting planet around the T Tauri star PTFO 8-8695, also known as CVSO 30, is essentially equivalent to the stellar rotation period (van Eyken et al. 2012; Ciardi et al. 2015; Johns-Krull et al. 2016b). Johns-Krull et al. (2016a) found a period ratio of $P_{orb}/P_{rot} \approx 1.23$ for the candidate $\sim$11–12 $M_{Jup}$ planet around the T Tauri star CI Tau. Lastly, the planet V830 Tau b has a period ratio of $P_{orb}/P_{rot} \approx 1.8$ (Donati et al. 2016).

### 7.11. *Uniqueness of RIK-210*

Inspection of more than 1400 *K2* light curves of secure or likely members of Upper Sco has revealed twelve sources with persistent or transient, short-duration flux dips that in some cases are apparently in phase with the stellar rotation (Stauffer et al. 2017, submitted). It is possible that RIK-210 may belong to this newly discovered class of stars, but we emphasize that RIK-210 is an outlier from these stars in nearly every regard. Namely, it has the longest rotation period by more than a factor of two, it is two spectral classes earlier in type than any of the other mentioned stars, and it has by far the deepest and most variable dimming events. Furthermore, none of the above-mentioned stars have demonstrated shallow, short-duration flux dips that are seemingly out of phase, or drifting in phase with respect to the stellar rotation.

Nevertheless, the physical scenarios explored in this work, and their presumed likelihoods, apply equally well to the population of variables identified in (Stauffer et al. 2017, submitted). Namely, those stars also lack primordial protoplanetary disks or spectroscopic accretion indicators. Those authors conclude the most likely culprits for the variability is material trapped in the magnetosphere or collisional debris.

### 8. CONCLUSIONS

We present evidence for transiting gas, dust, or debris around the 5–10 Myr old star RIK-210. Speculative sources of the obscuring material could be a magnetospheric cloud, an accretion flow from residual (yet undetected) gas and dust, remnants of the late stages of planet formation, the product of a giant-impact type collision, an enshrouded protoplanet with an extended tail, or one or more eccentric bodies undergoing periodic tidal disruption upon each periastron passage.

The most important aspect of the dimming events, which must be explained by any successful theory, is that they appear in phase with the stellar rotation. Theories invoking material tied to magnetic field lines at the corotation radius may explain this behavior, but do not readily account for the sub-percent dimmings seen elsewhere in the light curve that are not in phase with the stellar rotation.

The dimming events are variable in depth, duration, and morphology. The depths, while deep ($\sim$5–20%), can be produced by only a modest amount of dust. The durations are long ($\sim$6–18 hours), though the minimum duration is only somewhat longer than the expected transit duration at the corotation radius. Nevertheless, the lengthy durations imply a large size for the transiting cloud relative to the star. It is also possible that the occulting material is distributed in a torus which is tilted with respect to our line-of-sight, thereby producing dimmings only when a vertically extended part of the torus crosses the star.

Curiously, archival time series photometry from WASP provide no clear evidence for dimming events in the past. While some $\sim$0.1–0.2 mag dimmings are present in previous years they are not at a consistent phase and they are often the result of only one to two nights of data in a given year. We can not conclusively ascribe these dimmings to the same events observed by *K2*. Furthermore, follow-up photometry from LCOGT indicates the dimming events are no longer occurring at the level and phase expected from *K2*. However, our phase coverage in follow-up photometry is rather incomplete. Nevertheless, it is clear the transit signatures are transient in nature.

Follow-up radial velocity monitoring, a secondary spectral line search, and high-resolution imaging place stringent limits on the presence of any putative companions. RIK-210 is an apparently single star, and the upper limit of a companion orbiting at the corotation radius is $\sim$8 $M_{Jup}$. We note RV variability at the level of $\sim$2 km s$^{-1}$, the dominant component of which is likely induced by starspots. When phased to the *K2* ephemeris, the RVs show small point-to-point scatter in phase but do not exhibit the sinusoidal variation expected for spot-induced variability. It is possible that orbital motion due to one or more companions may contribute to the RV variability.

RIK-210 is diskless, as implied by its SED and a lack of spectroscopic accretion indicators. A modest amount of dust may remain at AU scales, from a weak 22 $\mu$m excess.

Close inspection of spectral line profiles reveal a chromospherically active star. No spectroscopic peculiarities are observed at the predicted phase of the dimming events, or elsewhere, though it is not clear the dimming events were still occurring at the time of our spectroscopic observations.

Continued photometric monitoring is needed to ascertain whether the dimmings observed by $K2$ have changed depth. High precision photometry is necessary in order to also detect the small, $\lesssim 1\%$ dips. These small dips may hold the clue to the correct physical interpretation of the larger-scale variability. We advocate for continuous or semi-continuous photometric monitoring, not just at the predicted phase of the primary dip.

If the dimming events reappear, or evaded our detection in follow-up photometry, a number of experiments may be conducted to clarify their physical origin. Observations of the Rossiter-McLaughlin effect would conclusively determine whether the flux decrements in RIK-210 are due to material in orbit around the star. For the depths observed by $K2$ (though now presumed to be shallower or entirely absent) the expected R-M amplitude is on the order of $1\,\mathrm{km\,s^{-1}}$. Multi-band photometric monitoring can be used to test whether the dip depths are wavelength-dependent; solid-body transits are achromatic, while extinction by dust is less severe at redder wavelengths. Finally, spectroscopic monitoring while the star is known to be dimming can test whether there is enhanced absorption by a gaseous cloud transiting the star.

The authors thank Todd Boroson for allocation of LCOGT director's discretionary time and Nikolaus Volgenau for assistance scheduling observations. We thank the anonymous referee for a thorough review, Saul Rappaport for helpful comments on an early draft and for providing the Fisher matrix analysis of the RVs, Jim Fuller for bringing $\sigma$ Ori E to our attention, Eugene Chiang, Konstantin Batygin, Kat Deck, Brad Hansen, and Lee Hartmann for helpful discussions, as well as Norio Narita and John Livingston for attempting follow-up observations. T.J.D. is supported by an NSF Graduate Research Fellowship under Grant DGE1144469. E.A.P. is supported through a Hubble Fellowship. A.M.C.'s research was supported by an appointment to the NASA Postdoctoral Program at the NASA Ames Research Center, administered by Universities Space Research Association under contract with NASA. B.J.F. was supported by the National Science Foundation Graduate Research Fellowship under grant No. 2014184874. Any opinion, findings, and conclusions or recommendations expressed in this material are those of the authors and do not necessarily reflect the views of the National Science Foundation. This paper includes data collected by the $Kepler/K2$ mission, funded by the NASA Science Mission directorate and obtained from the Mikulski Archive for Space Telescopes (MAST), supported by the NASA Office of Space Science via grant NNX09AF08G. Some of the data presented herein were obtained at the W.M. Keck Observatory, which is operated as a scientific partnership among the California Institute of Technology, the University of California and the National Aeronautics and Space Administration. The Observatory was made possible by the generous financial support of the W.M. Keck Foundation. The authors wish to recognize and acknowledge the very significant cultural role and reverence that the summit of Mauna Kea has always had within the indigenous Hawaiian community. We are most fortunate to have the opportunity to conduct observations from this mountain.